\documentclass{article}

\usepackage[preprint, nonatbib]{neurips_2019}
\usepackage[utf8]{inputenc} 
\usepackage[T1]{fontenc}    
\usepackage{hyperref}       
\usepackage{url}            
\usepackage{booktabs}       
\usepackage{amsfonts}       
\usepackage{nicefrac}       
\usepackage{microtype}      
\usepackage{mdframed,lipsum,calc,tikz}

\usepackage{amsmath}
\usepackage{amssymb}
\usepackage{graphicx}
\usepackage{booktabs}
\usepackage{caption}
\usepackage{subcaption}
\usepackage{multirow}
\usepackage{xcolor}
\usepackage{xspace}
\usepackage{todonotes}
\usepackage{url}            
\usepackage[shortlabels]{enumitem}
\usepackage{wrapfig}
\usepackage{algorithm}
\usepackage{algpseudocode}
\usepackage{tabularx}
\usepackage{listings}
\usepackage{xcolor}

\newcommand{\sys}[0]{Mooncake\xspace}

\lstdefinestyle{jsonStyle}{
    basicstyle=\small\ttfamily,
    columns=fullflexible,
    showstringspaces=false,
    commentstyle=\color{gray}\itshape,
    stringstyle=\color{blue},
    morestring=[b]",
    morecomment=[l]{//},
    morecomment=[s]{/*}{*/},
    frame=single,
    framerule=0pt,
    backgroundcolor=\color{gray!10},
    rulecolor=\color{black!30},
    framesep=5pt,
    aboveskip=10pt,
    belowskip=10pt
}

\makeatletter

\newif\ifarxiv
\arxivfalse

\begin{document}

\title{Mooncake: A KVCache-centric Disaggregated Architecture for LLM Serving}

\author{
Ruoyu Qin$^{\spadesuit\heartsuit1}$~~~~Zheming Li$^{\spadesuit1}$~~~~Weiran He$^{\spadesuit}$ \\
\And
Mingxing Zhang$^{\heartsuit2}$~~~~Yongwei Wu$^{\heartsuit}$~~~~Weimin Zheng$^{\heartsuit}$~~~~Xinran Xu$^{\spadesuit2}$ \\
$^{\spadesuit}$Moonshot AI~\medskip~$^{\heartsuit}$Tsinghua University}
\footnotetext[1]{~Ruoyu Qin's part of work done as an intern at Moonshot AI, contributed equally with Zheming Li.}\footnotetext[2]{~Corresponding to zhang\_mingxing@mail.tsinghua.edu.cn,~xuxinran@moonshot.ai.}

\maketitle

\begin{abstract}

Mooncake is the serving platform for Kimi, a leading LLM service provided by Moonshot AI. It features a KVCache-centric disaggregated architecture that separates the prefill and decoding clusters. It also leverages the underutilized CPU, DRAM, and SSD resources of the GPU cluster to implement a disaggregated cache of KVCache. The core of Mooncake is its KVCache-centric scheduler, which balances maximizing overall effective throughput while meeting latency-related Service Level Objectives (SLOs). Unlike traditional studies that assume all requests will be processed, Mooncake faces challenges due to highly overloaded scenarios. To mitigate these, we developed a prediction-based early rejection policy. Experiments show that Mooncake excels in long-context scenarios. Compared to the baseline method, Mooncake can achieve up to a 525\% increase in throughput in certain simulated scenarios while adhering to SLOs. Under real workloads, Mooncake's innovative architecture enables Kimi to handle 75\% more requests.

\end{abstract}

\section{Introduction}
\subsection{Motivation of Developing Mooncacke}

With the rapid adoption of large language models (LLMs) in various scenarios~\cite{chatgpt, llama2, chen2021evaluating,memgpt}, the workloads for LLM serving have become significantly diversified. These workloads differ in input/output length, frequency and distribution of arrival, and, most importantly, demand different kinds of Service Level Objectives (SLOs). As a Model as a Service (MaaS) provider, one of the primary goals of Kimi~\cite{kimi} is to solve an optimization problem with multiple complex constraints. The optimization goal is to maximize overall effective throughput, which directly impacts revenue, while the constraints reflect varying levels of SLOs. These SLOs typically involve meeting latency-related requirements, mainly the time to first token (TTFT) and the time between tokens (TBT).

\begin{figure*}[th]
\begin{center}
\includegraphics[width=\textwidth]{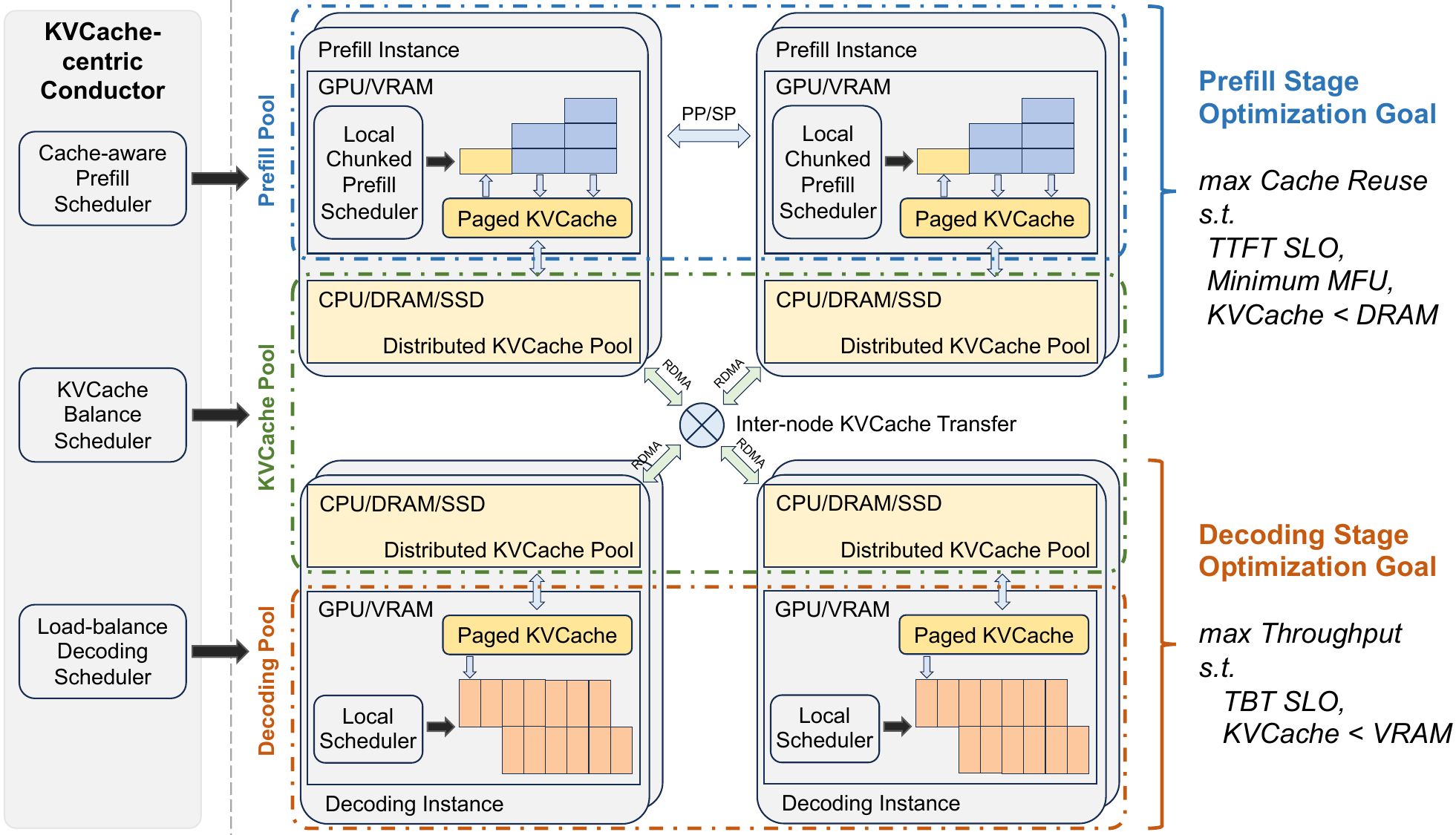}
\caption{Mooncake Architecture.}
\label{fig:arch}
\end{center}
\end{figure*}

To achieve this goal, a prerequisite is to make the best use of the various kinds of resources available in the GPU cluster. 
Specifically, although GPU servers are currently provided as highly integrated nodes (e.g., DGX/HGX supercomputers~\cite{nvidia-h100}), it is necessary to decouple and restructure them into several disaggregated resource pools, each optimized for different but collaborative goals. For example, many researchers~\cite{patel2023splitwise, zhong2024distserve, hu2024inference} have suggested separating prefill servers from decoding servers because these two stages of LLM serving have very different computational characteristics, in which the KVCache shifts with requests moving from prefill to decoding servers. 

Building on this idea, we found that the scheduling of KVCache is central to LLM serving scheduling. To improve overall throughput, there are typically two general approaches: {\em 1)} reuse KVCache as much as possible to reduce the required computation resources; and {\em 2)} maximize the number of tokens in each batch to improve the Model FLOPs Utilization (MFU). However, reusing KVCache from a remote location will prolong the TTFT, and a large batch size will lead to a larger TBT. Thus, the utilization of both these throughput-oriented optimizations may lead to violations of latency-related SLOs.

According to the above guidelines, we propose a disaggregated design that is centered around KVCache for scheduling and optimization. Figure~\ref{fig:arch} presents our current {\bf KVCache-centric disaggregated architecture} for LLM serving, named Mooncake. For each request, the global scheduler (Conductor) needs to select a pair of prefill and decoding instances and schedule the request in the following steps: {\em 1)} transfer as much reusable KVCache as possible to the selected prefill instance; {\em 2)} complete the prefill stage in chunks/layers and continuously stream the output KVCache to the corresponding decoding instance; {\em 3)} load the KVCache and add the request to the continuous batching process at the decoding instance for generating request outputs.

Although this process seems straightforward, the selection policy is complex due to many restrictions. In the prefill stage, the main objective is to reuse the KVCache as much as possible to avoid redundant computation. However, waiting for KVCache stored on lower-tier storage may violate the TTFT SLO. Additionally, high demand on the KVCache server can lead to network congestion, prolonging the waiting time. 
Thus Conductor is also responsible for predicting the future usage of KVCache blocks and executing scheduling operations such as swapping and replication accordingly. 
The hottest blocks should be replicated to multiple nodes to avoid fetching congestion, while the coldest ones should be swapped out to reduce reserving costs.
Prefill scheduling is also constrained by the availability of DRAM space in the prefill node, especially when much of the memory is reserved for the global KVCache pool.

In contrast, the decoding stage has different optimization goals and constraints. The aim is to aggregate as many tokens as possible in a decoding batch to improve MFU. 
However, this objective is restricted not only by the TBT SLO but also by the total size of the aggregated KVCache that can be contained in the VRAM.

More importantly, existing research on LLM serving assumes sufficient resources and focuses on improving resource utilization. In contrast, the current GPU/accelerator supply is limited, and many MaaS providers face severe overload problems, especially during peak times. Scheduling in such scenarios presents unique challenges that existing works have not explored. For example, we need to predict future loads and reject certain requests early if there will be no available decoding slots after the prefill stage, to save wasted computation resources.
However, a straightforward implementation of such an early reject policy surprisingly leads to fluctuations in the overloads. This has led us to aim at predicting the generation length of specific queries and making overall load predictions in the short-term future to implement a better rejection policy. It is also necessary to classify different request priorities to implement priority-based scheduling. 
In this paper, we summarize these problems as {\bf overload-oriented scheduling} and present our preliminary study results.

\subsection{Design and Results of Mooncacke}

In the following sections of this paper, we first present an overview of Mooncake's architecture, including its main components and the typical workflow for processing a request (\S\ref{sec:overview}). Then, we describe the main design choices made during its implementation, especially those not covered in current research.

First, in \S\ref{sec:prefill}, we discuss how to implement a separate prefill node pool that seamlessly handles the dynamic distribution of context length. We employ a chunked pipeline parallelism (CPP) mechanism to scale the processing of a single request across multiple nodes, which is necessary for reducing the TTFT of long-context inputs. Compared to traditional sequence parallelism (SP) based solutions, CPP reduces network consumption and simplifies the reliance on frequent elastic scaling. This mechanism is further supplemented with layer-wise prefill that enables stream transferring of KVCache to overlap latency.

Next, in \S\ref{section:ccs}, we detail our KVCache-centric request scheduling algorithm, which balances instance loads and user experience as measured by TTFT and TBT SLOs. This includes a heuristic-based automated hot-spot migration scheme that replicates hot KVCache blocks without requiring precise predictions of future KVCache usage. Experimental results show that our cache-aware scheduling can significantly lower TTFT in real-world scenarios. In end-to-end experiments using public datasets, simulated data, and real workloads, Mooncake excels in long-context scenarios. Compared to the baseline method, Mooncake can achieve up to a 525\% increase in throughput while meeting SLOs. Under real workloads, Mooncake enables Kimi to handle 75\% more requests.

Finally, unlike existing work on LLM serving that assumes all requests will be processed, Mooncake consistently faces overload due to Kimi's rapid growth in user requests. Thus, Mooncake's scheduling involves determining whether to accept or reject incoming requests based on the system load. In \S\ref{sec:oos}, we discuss our implementation of a unique early rejection policy that reduces wasted computational resources in overloaded scenarios. We further explore the load fluctuation problem caused by straightforward early rejection and how predicting future load can mitigate this issue.

Mooncake is currently the primary platform for serving Kimi and has successfully handled exponential workload growth, proving its effectiveness in scaling out to large and highly overloaded workloads. However, many more problems need to be explored, and these future directions are also included in the paper.

\global\mdfdefinestyle{exampledefault}{%
linecolor=black,linewidth=3pt,%
leftmargin=1cm,rightmargin=1cm
}
\begin{mdframed}[style=exampledefault]
To protect proprietary information and facilitate reproducibility, all the experimental results reported in this paper are based on replayed traces of real workloads, but using a {\bf dummy model} that follows the same architecture as LLaMA2-70B. The trace includes only the timing of request arrivals, the number of input tokens, and the number of output tokens, the remaped block hash, without any real user content. The trace is open-sourced at \url{https://github.com/kvcache-ai/Mooncake}.
\end{mdframed}

\section{Preliminary and Problem Definition}
\label{sec:preliminary}

Modern large language models (LLMs) are based on the Transformer architecture, which utilizes attention mechanisms and multilayer perceptrons (MLPs) to process input. Popular Transformer-based models, such as GPT~\cite{radford2019language} and LLaMA~\cite{touvron2023llama}, employ a decoder-only structure. Each inference request is logically divided into two stages: the prefill stage and the decoding stage.

\begin{figure*}[th]
\begin{center}
\includegraphics[width=\textwidth]{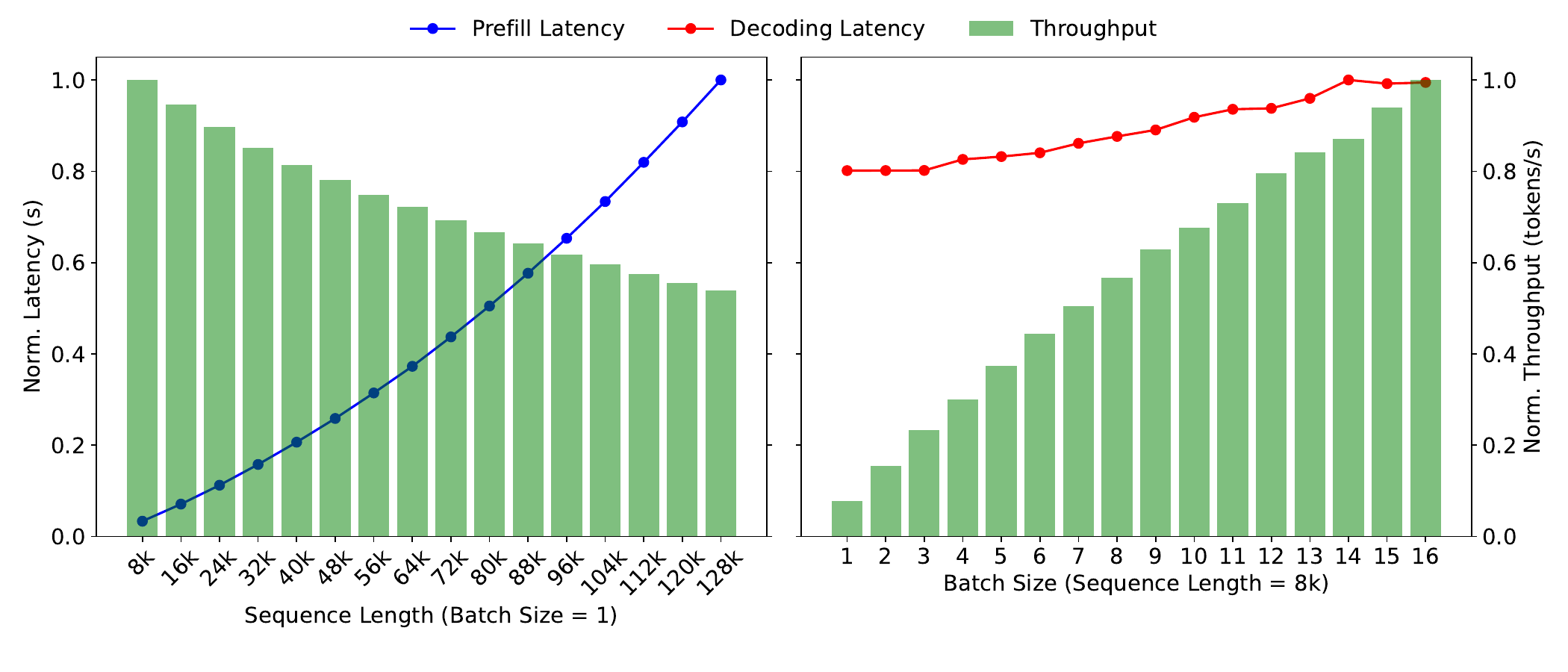}
\caption{Normalized throughput and latency of prefill and decoding stages with different sequence lengths or batch sizes for the dummy 
LLaMA2-70B model.}
\label{fig:prefill_decode_latency}
\end{center}
\end{figure*}

In the prefill stage, all input tokens are processed in parallel. This stage generates the first output token while storing intermediate results of computed keys and values, referred to as the KVCache. The decoding stage then uses this KVCache to autoregressively generate new tokens, adding new keys and values from the computation to the KVCache. The ability to process input tokens simultaneously in the prefill stage typically makes it computationally intensive, except for short requests. Since the computational complexity of attention networks scales quadratically with input length while the complexity of MLP scales linearly, computation time in the prefill stage generally increases superlinearly with input length, as shown in the left part of Figure~\ref{fig:prefill_decode_latency}.

In contrast, the decoding stage processes only one token at a time per batch due to the limitation of autoregressive generation. This makes it memory-constrained and causes computation time to increase sublinearly with batch size, as shown in the right part of Figure~\ref{fig:prefill_decode_latency}. A widely used optimization in the decoding stage is continuous batching~\cite{yu2022orca, kwon2023efficient}. Before each iteration, the scheduler checks the status of all requests, adding newly arrived requests to the batch's prefill stage while removing completed requests.

Due to the distinct characteristics of the prefill and decoding stages, MaaS providers set different metrics to measure their corresponding Service Level Objectives (SLOs). Specifically, the prefill stage is mainly concerned with the latency between the request arrival and the generation of the first token, known as the time to first token (TTFT). On the other hand, the decoding stage focuses on the latency between successive token generations for the same request, referred to as the time between tokens (TBT).

As a MaaS provider, it is crucial to ensure quality assurance by meeting SLO metrics defined by service agreements. 
For example, a metric such as $\text{TTFT}_{P90} = 4\times$ indicates that 90\% of inference requests will have a TTFT no greater than four times that of a single request running under the same conditions without interference. Specifically, in the end-to-end experiment of this paper (\S\ref{sec:end_to_end_performance}), we set $\text{TTFT}_{P90} = 10\times$ and $\text{TBT}_{P90} = 5\times$. In real deployments, we set fixed SLOs of TTFT and TBT. If monitoring detects unmet SLOs, we either add inference resources or reject some incoming requests.

However, due to the current contingent supply of GPUs, elastically scaling out the inference cluster is typically unfeasible. Therefore, deciding which requests to reject becomes a core issue in overload-oriented scheduling. Our main objective is to maximize overall throughput while adhering to SLOs, a concept referred to as goodput in other research~\cite{zhong2024distserve, wu2024loongserve}.
Our approach differs in that only requests that fully complete their execution are counted in the measure of goodput. Otherwise, all previously consumed/generated tokens are not counted, and the corresponding resources are wasted. In other words, a request should be rejected as early as possible if it cannot finish its full execution under the SLO. Achieving this goal involves not only optimizing the architecture of both the prefill and decoding stages but also developing a capability to predict short-term future loads.

\section{Overview of Mooncake's Disaggregated Architecture}\label{sec:overview}

As depicted in Figure \ref{fig:arch}, Mooncake employs a disaggregated architecture that not only separates prefill from decoding nodes, but also groups the CPU, DRAM, SSD, and RDMA resources of the GPU cluster to implement a disaggregated KVCache. 
This disaggregated cache harnesses underutilized resources to provide ample cache capacity and transfer bandwidth, enabling efficient near-GPU prefix caching without additional costs.

\begin{figure*}[th]
\begin{center}
\includegraphics[width=\textwidth]{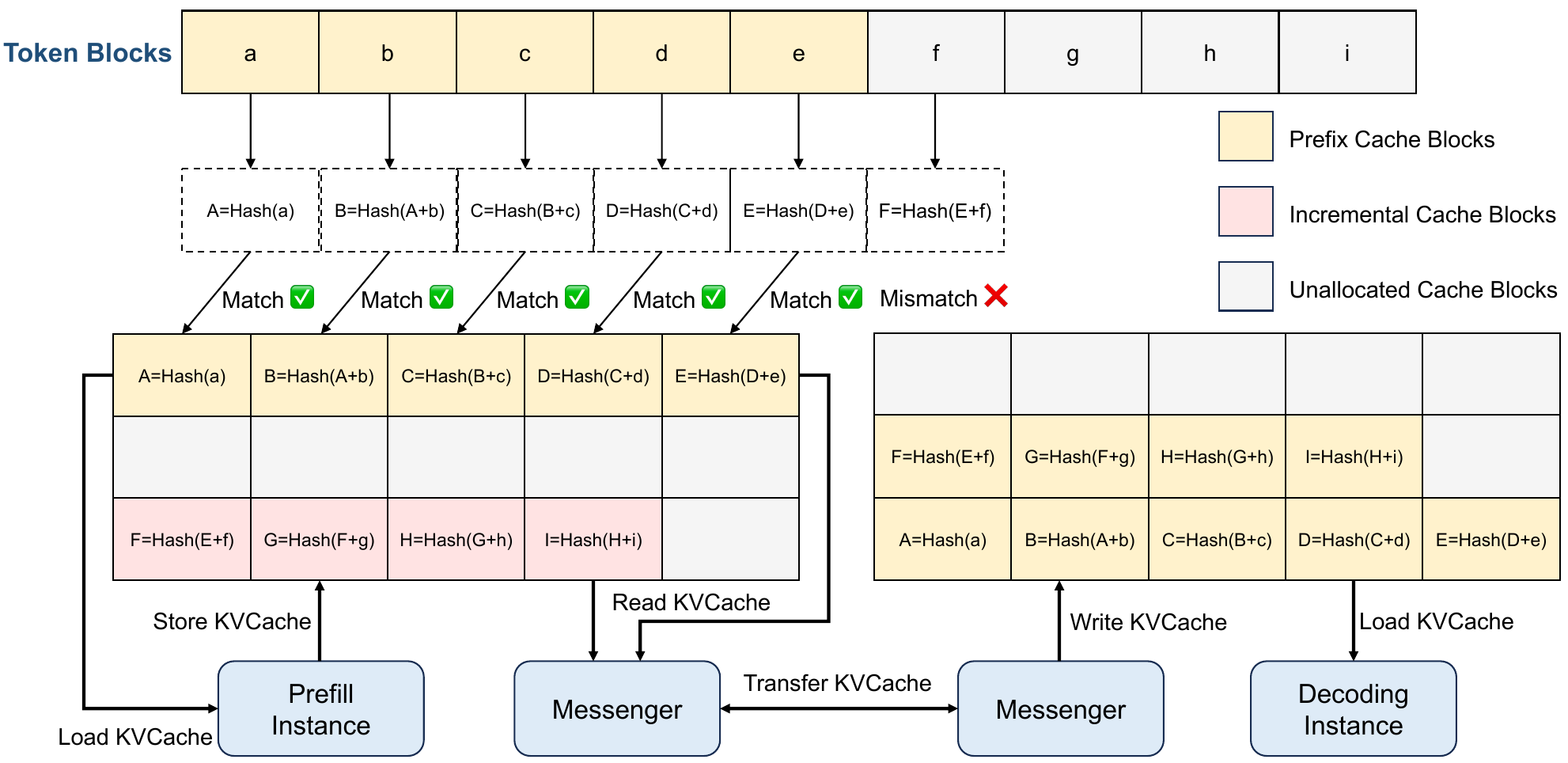}
\caption{The KVCache pool in CPU memory. Each block is attached with a hash value determined by both its own hash and its prefix for deduplication.}
\label{fig:cache_unit}
\end{center}
\end{figure*}

Figure \ref{fig:cache_unit} illustrates the storage and transfer logic of the KVCache blocks. In CPU memory, KVCache is stored as paged blocks. Depending on the request patterns, it can use cache eviction algorithms such as LRU (Least Recently Used), LFU (Least Frequently Used), or algorithms based on request characteristics. The transfer of these KVCache blocks across CPUs and GPUs is handled by a separate (GPUDirect) RDMA-based component called Messenger.
This architecture also enables us to provide the context caching API to outside users for a higher reuse of KVCache.

To schedule all these disaggregated components, at its center, Mooncake implements a global scheduler named Conductor. 
Conductor is responsible for dispatching requests based on the current distribution of the KVCache and workloads. It also replicates or swaps certain blocks of the KVCache if it is beneficial for future inference.
Specifically, Figure \ref{fig:inference_instance_workflow} demonstrates the typical workflow of a request. Once tokenizing is finished, the conductor selects a pair of prefill nodes and a decoding node, and starts a workflow comprising four steps:

\begin{figure*}[th]
\begin{center}
\includegraphics[width=\textwidth]{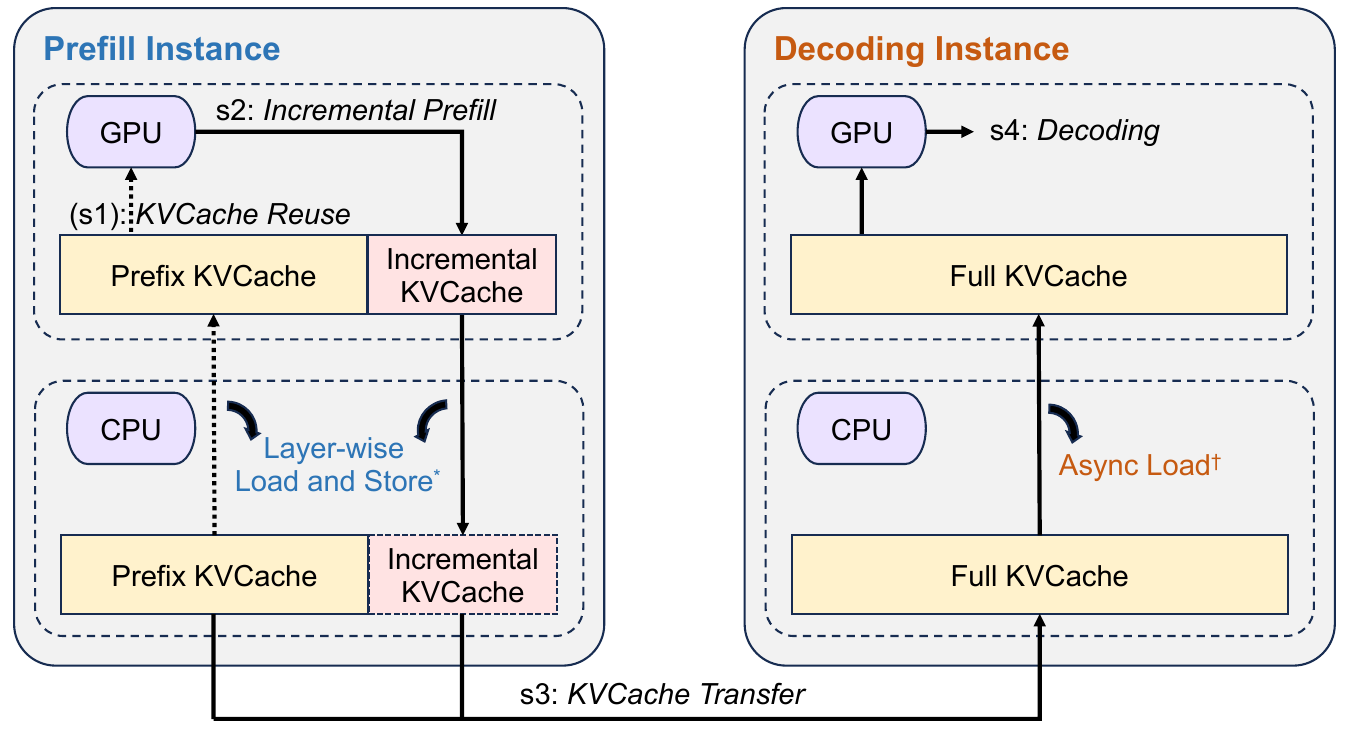}
\caption{Workflow of inference instances. ($*$) For prefill instances, the load and store operations of the KVCache layer are performed layer-by-layer and in parallel with the prefill computation to mitigate transmission overhead (see \S\ref{sec:layerwise-prefill}).  ($\dagger$) For decoding instances, asynchronous loading is performed concurrently with GPU decoding to prevent GPU idle time.}
\label{fig:inference_instance_workflow}
\end{center}
\end{figure*}

\noindent\textit{\underline{1) KVCache Reuse:}} The selected prefill node (group) receives a request that includes the raw input, the block IDs of the prefix cache that can be reused, and the block IDs of the full cache allocated to the request. It loads the prefix cache from remote CPU memory into GPU memory based on the prefix cache block IDs to bootstrap the request. This step is skipped if no prefix cache exists. This selection balances three objectives: reusing as much KVCache as possible, balancing the workloads of different prefill nodes, and guaranteeing the TTFT SLO.
It leads to a KVCache-centric scheduling that will be further discussed in \S\ref{section:ccs}.

\noindent\textit{\underline{2) Incremental Prefill:}} The prefill node (group) completes the prefill stage using the prefix cache and stores the newly generated incremental KVCache back into CPU memory. If the number of uncached input tokens exceeds a certain threshold ($prefill\_chunk$), the prefill stage is split into multiple chunks and executed in a pipeline manner. This threshold is selected to fully utilize the corresponding GPU's computational power and is typically larger than 1000 tokens. The reason for using chunked but still disaggregated prefill nodes is explained in \S\ref{sec:prefillpp}.

\noindent\textit{\underline{3) KVCache Transfer:}} The aforementioned Messenger service is deployed in each node to manage and transfer these caches. Each Messenger operates as an independent process within its respective inference instance, receiving signals to facilitate high-speed, cross-machine KVCache transfer. This step is asynchronously executed and overlapped with the above incremental prefill step, streaming the KVCache generated by each model layer to the destination decoding node's CPU memory to reduce waiting time.

\noindent\textit{\underline{4) Decoding:}} After all the KVCache is received in the CPU DRAM of the decoding node, the request joins the next batch in a continuous batching manner. Conductor pre-selects the decoding node based on its current load to ensure it does not violate the TBT SLO. However, this SLO is double-checked by the local scheduler because the anticipated load may have changed after the prefill stage. This double-checking may lead to the rejection of the request, in which case the corresponding prefill costs are wasted.

\begin{figure*}[th]
\centering
\includegraphics[width=0.9\linewidth]{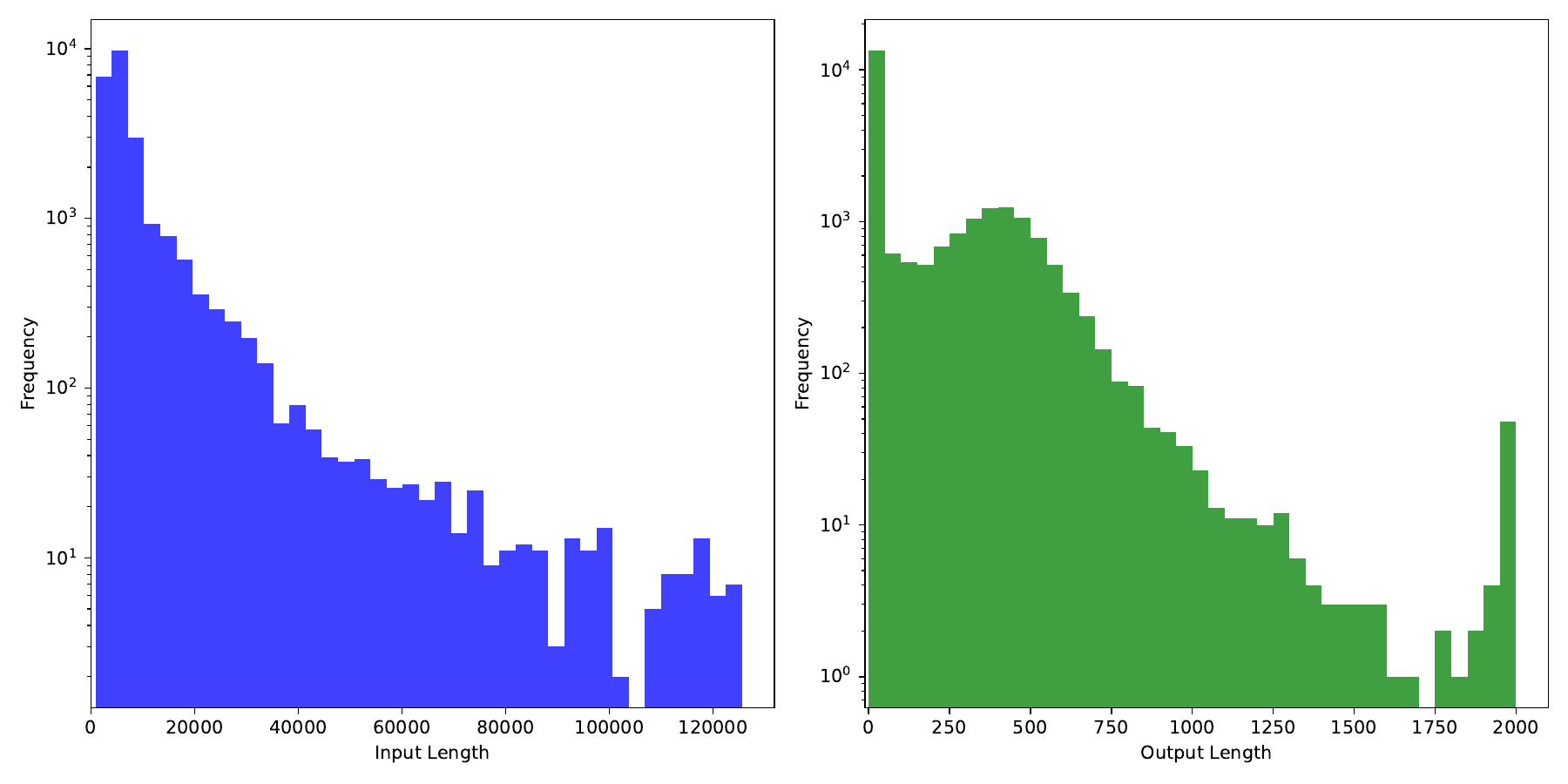}
\caption{Input and output length distributions in the request trace.}
\label{fig:trace_length_input_output}
\end{figure*}

\section{Sampled Real-world Request Trace}

To facilitate further research on LLM serving, we sampled a subset of online request data from a 1-hour period to create an open-source request trace. 
To preserve caching relationships between requests, we prioritized collecting requests within the same session. 
The trace dataset comprises 23,608 entries, with fields including \textit{timestamp}, \textit{input\_length}, \textit{output\_length}, and \textit{hash\_ids}. 
We included the remapped block hashes, which are particularly useful for analyzing and implementing KVCache reuse policies. 
To the best of our knowledge, this is the first open-source dataset that can be used for real-world reuse analysis.

\subsection{Data Details}

\begin{lstlisting}[style=jsonStyle, frame=shadowbox, caption={Request samples.}, label={lst:trace_samples}]
{
    "timestamp": 27482,
    "input_length": 6955,
    "output_length": 52,
    "hash_ids": [46, 47, 48, 49, 50, 51, 52, 53, 54, 55, 56, 57, 2353, 2354]
}
{
    "timestamp": 30535,
    "input_length": 6472,
    "output_length": 26,
    "hash_ids": [46, 47, 48, 49, 50, 51, 52, 53, 54, 55, 56, 57, 2366]
}
\end{lstlisting}

Listing~\ref{lst:trace_samples} presents two samples from our trace dataset. To protect our customers' privacy, we applied several mechanisms to remove user-related information while preserving the dataset's utility for simulated evaluation. The meanings of the fields are explained below.

\textbf{Timestamp:} The \textit{timestamp} field indicates the relative arrival times of requests, ranging from 0 to 3,600,000, in milliseconds.

\textbf{Input \& Output Length:} For privacy protection, our trace does not include actual text or tokens. Instead, it uses \textit{input\_length} and \textit{output\_length}, representing the number of input and output tokens, similar to Splitwise~\cite{patel2023splitwise}.

\begin{wrapfigure}{r}{0.42\textwidth}
\centering
\includegraphics[width=0.41\textwidth]{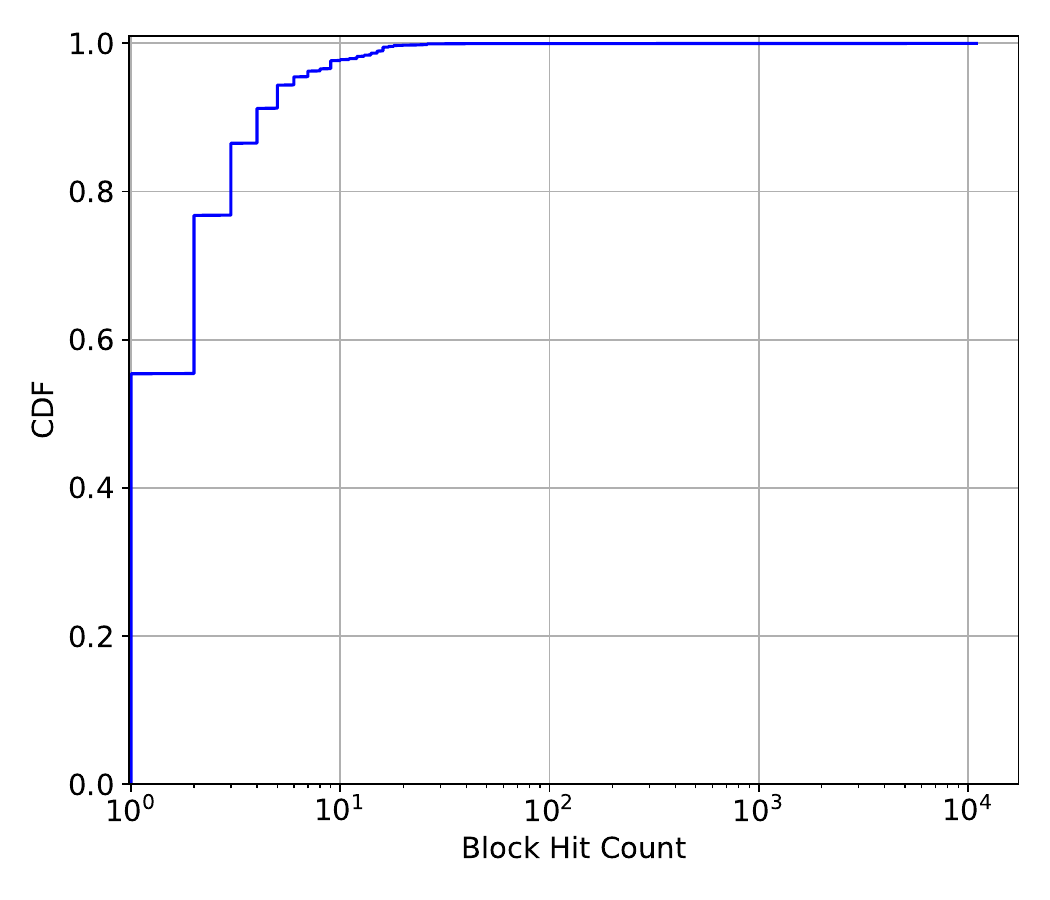}
\caption{CDF (Cumulative Distribution Function) of the block hit count in the request trace.}
\label{fig:block_hit_count}
\end{wrapfigure}  

\textbf{Hash ID:} The \textit{hash\_ids} field describes prefix caching relationships. It is generated by hashing token blocks (with a block size of 512) into prefix hash values that include both the current and all preceding blocks (detailed in Figure~\ref{fig:cache_unit}). The resulting hash values are then mapped to globally unique IDs. Identical hash IDs indicate that a block of tokens, along with preceding tokens, is the same, thus allowing reuse within the corresponding KVCache. For example, in the provided samples, the first 12 hash IDs are identical, indicating they can share prefix caching for the first 12*512=6,144 tokens.





\subsection{Statistical Features}




Figure~\ref{fig:trace_length_input_output} illustrates the distribution of input and output lengths in our trace, with an average input length of 8,590 tokens and an average output length of 182 tokens. The average input-output ratio is approximately 720. It is important to note that this is only a representative pattern and not unanimous for all workloads, reflecting Kimi's renowned capability for superior long-context processing and understanding.

We also conducted a simple cache policy analysis based on this trace, assuming a single global cache pool. Table~\ref{tab:cache_policy_capacity} compares three cache strategies: LRU, LFU, and LengthAwareCache (similar to LFU but prioritizing cache blocks occurring later in requests) across different cache capacities. Increasing the cache capacity from 1,000 to 50,000 blocks boosts the cache hit ratio from 34\% to 55\%. Further capacity increases show minimal improvement. However, this {\bf should not} be interpreted as an indication that larger caches are unnecessary, as the sample trace represents only a subset of real-world workloads. The required capacity should scale proportionally in actual scenarios. LRUCache performs best under this dataset's patterns, likely due to the temporal proximity in request utilization.  

Additionally, we observed a notable imbalance in cache block popularity, with over 50\% of cache blocks remaining unused while certain blocks are accessed tens of thousands of times, as shown in Figure~\ref{fig:block_hit_count}. Replicating these hot blocks is essential to avoid transfer congestion.

\begin{table*}[th]
\begin{small}
\begin{center}
\caption{Cache hit rates under different cache policies and capacities.}
\label{tab:cache_policy_capacity}
\begin{tabular}{ccccccc}
\toprule
Block capacity & Inf & 100000 & 50000 & 30000 & 10000 & 1000 \\
\midrule
LRUCache & 0.55 & 0.55 & 0.55 & 0.54 & 0.46 & 0.34 \\
LFUCache & 0.55 & 0.55 & 0.55 & 0.50 & 0.40 & 0.34 \\
LengthAwareCache & 0.55 & 0.54 & 0.53 & 0.50 & 0.40 & 0.34 \\
\bottomrule
\end{tabular}
\end{center}
\end{small}
\end{table*}



\section{Implementation of the Prefill Pool}\label{sec:prefill}

Unlike the inviolable decoding nodes, the necessity and best practices for designing a separate and elastic prefill pool remain under debate.
For example, although many researchers~\cite{patel2023splitwise, zhong2024distserve, hu2024inference} share our intuition to use a disaggregated architecture, it is worth discussing whether this separation is still necessary with the introduction of chunked prefill~\cite{agrawal2024taming}. 
Chunked prefill divides the input tokens into multiple small chunks that join the continuous batch process. This approach has two clear benefits: 
{\em 1)} Without separation, all nodes are treated equally, making scheduling easier;
{\em 2)} Inlining chunked prefill into the decoding batch can improve the computational intensity of the decoding batch, leading to better MFU.

However, after careful consideration, we decided to maintain Mooncake's disaggregated architecture. A request's prefill is inlined into the decoding batch only when it can be forwarded without chunking and without compromising the TBT SLO. 
There are two main reasons for this decision:
{\em 1)} Prefill nodes require different cross-node parallelism settings to handle long contexts (\S\ref{sec:prefillpp}).
{\em 2)} It presents a unique opportunity to save VRAM (\S\ref{sec:layerwise-prefill}).

\subsection{Multi-node Prefill}\label{sec:prefillpp}

The available context length of recent LLMs is increasing rapidly, from 8k to 128K and even 1M~\cite{gemini15pro}. 
Typically, for such long context requests, the input tokens can be 10 to 100 times larger than the output tokens, making optimizing the TTFT crucial. 
Due to the abundant parallelism in long context prefill, using more than a single 8x GPU node to process them in parallel is desirable. 
However, extending tensor parallelism  (TP) across more than one node requires two expensive RDMA-based all-reduce operations per layer, significantly reducing the MFU of prefill nodes.

Recently, many works have proposed sequence parallelism (SP)~\cite{jacobs2023deepspeed, liu2023ring, brandon2023striped, li2023lightseq, korthikanti2023reducing, li2023sequence, fang2024usp}. 
SP partitions the input sequences of requests across different nodes to achieve acceleration. 
These SP methods take advantage of the associative property of the attention operator and require cross-node communication at least once per layer during the implementation of Ring Attention~\cite{liu2023ring} or Striped Attention~\cite{brandon2023striped}. 
This greatly reduces network consumption and improves MFU.

However, adopting SP still results in a worse MFU compared to using single-node TP only. 
A desired deployment organizes prefill nodes into two groups: one with TP only and the other with SP. 
Requests are dispatched to the SP group only when necessary to meet the TTFT SLO. 
This further disaggregation leads to problems in dynamically adjusting the number of nodes in each group, as a static parallelism setting can result in low utilization across the cluster. 
Recent research~\cite{wu2024loongserve} proposes elastic sequence parallelism to dynamically scale up or down the SP group. 
Although possible, this adds complexity to our architecture. 
For example, it requires establishing a global communication group in advance and complicates Conductor's design when considering metrics like cache reuse utilization and SLO requirement violations during adjustments. 
This makes it challenging for our situations that require frequent on-the-fly scalability during deployment.
Additionally, SP still requires frequent cross-node communication, which lowers the MFU and competes with network resources for transferring KVCache across nodes.

To address this, Mooncake leverages the autoregressive property of decoder-only transformers and implements chunked pipeline parallelism (CPP) for long context prefill. 
We group every X nodes in the prefill cluster into a pipelined prefill node group. 
For each request, its input tokens are partitioned into chunks, each no longer than the $prefill\_chunk$. 
Different chunks of the same request can be processed simultaneously by different nodes, thus parallelizing the processing and reducing TTFT.

CPP offers two main benefits: {\em 1)} Similar to pipeline parallelism in training, it requires cross-node communication only at the boundaries of each pipeline stage, which can be easily overlapped with computation. This leads to better MFU and less network resource contention with KVCache transfer. {\em 2)} It naturally fits both short and long contexts, bringing no significant overhead for short context prefill and avoiding frequent dynamic adjustment of node partitioning. 
This pipeline-based acceleration method has been explored in training systems~\cite{li2021terapipe}, but to our knowledge, this is the first application in the inference stage, as long context inference has only recently emerged.

\subsection{Layer-wise Prefill}
\label{sec:layerwise-prefill}

Beyond computational power, the limited size of VRAM is also a precious resource, and we aim to minimize the VRAM occupation by states, primarily the KVCache. 
Theoretically, if the KVCache size of a request is $S$ and the processing time is $T$, its occupation cost is $S*T$. 
If a request is chunked and the processing of each chunk is inlined with other decoding requests in chunked prefill, $T$ will increase, leading to a larger occupation cost.

\begin{wrapfigure}{r}{0.5\textwidth}
\centering
\includegraphics[width=0.48\textwidth]{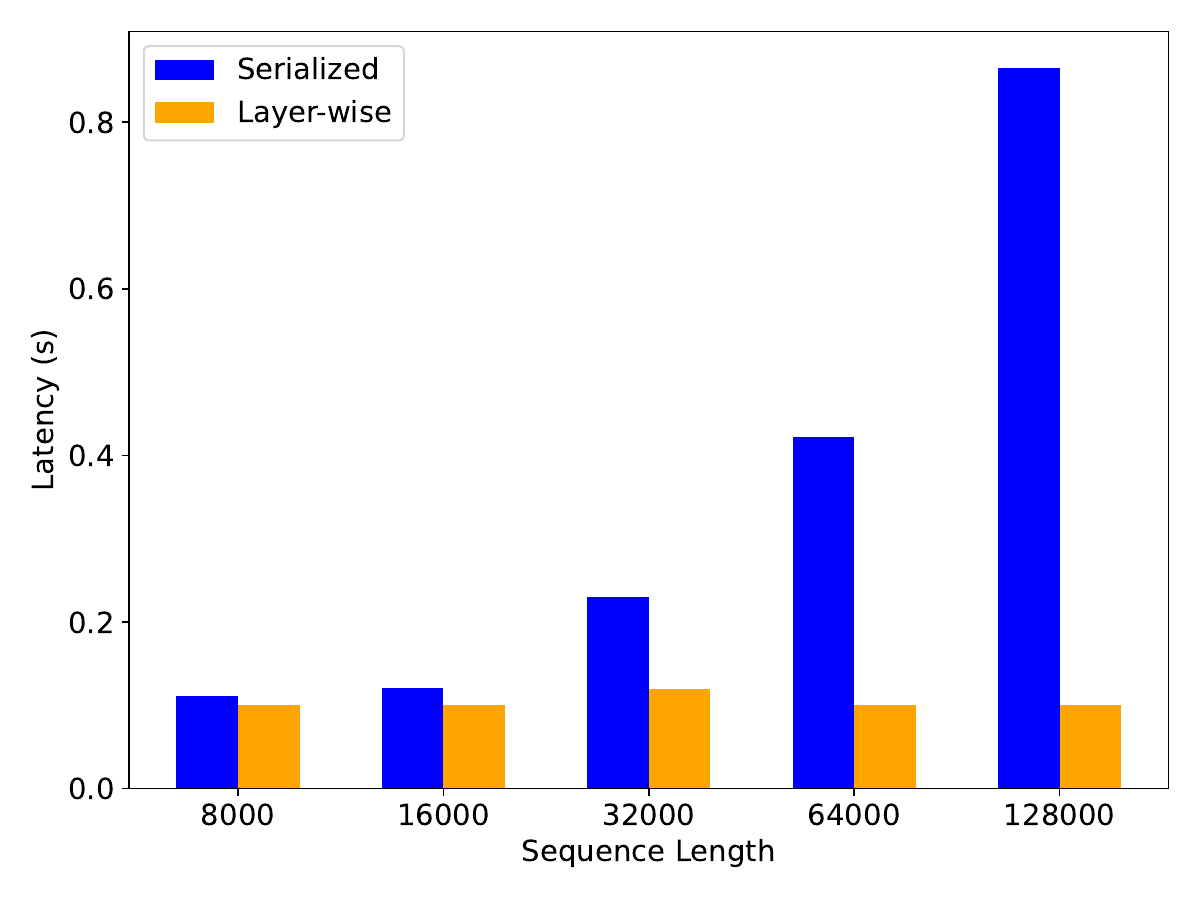}
\caption{Latency of storing KVCache of different request lengths (Layer-wise latency refers to the difference in latency between Layer-wise Prefill and Prefill without storing KVCache).}
\label{fig:overlap_latency}
\end{wrapfigure}

Moreover, since prefill is processed layer-by-layer and is computation-bound, it is possible to overlap the transferring and dumping of KVCache with computation, further reducing its occupation cost. 
In Mooncake, KVCache loading and storing are executed asynchronously via launch and wait operations. Before each layer's attention computation begins, the model waits for the asynchronous loading of that layer's KVCache to complete and triggers the next layer's asynchronous KVCache loading. After the attention calculation is complete, asynchronous storage of that layer's KVCache is launched. Once all layers' computations are finished, the process waits for the completion of all asynchronous storage operations. Transfer overlapping allows the prefill instance's execution time to be roughly equivalent to either the KVCache loading time or the standard prefilling time, depending on the prefix cache proportion relative to the input length. The experimental result of KVCache storing latency, as shown in Figure~\ref{fig:overlap_latency}, demonstrates that the layer-wise prefill can effectively reduce the latency for long-context requests.

The main advantage of this overlap effectiveness is that it enables us to disregard the available VRAM size in prefill scheduling, as long as it can contain a single request. 
As shown in Figure \ref{fig:arch}, the scheduling of prefill nodes only considers the KVCache distribution and the available DRAM size.

In the future, we intend to explore more uses for this free VRAM. For example, OpenAI recently proposed the use of batch APIs~\cite{openai2024batch}, which enable users to send asynchronous groups of requests at 50\% lower costs, but with only a clear 24-hour turnaround time. This service is ideal for processing jobs that do not require immediate responses. Since there is no stringent TBT for these batch requests, we can inline even the decoding stage of these requests into prefill processing for better MFU, if there is enough VRAM space to hold the corresponding KVCache.
\section{KVCache-centric Scheduling}
\label{section:ccs}
In this section, we mainly discuss how Conductor schedules the requests and KVCache blocks under normal conditions, leaving the discussion on overload scenarios for the next section.

\algrenewcommand\algorithmicrequire{\textbf{Input:}}
\algrenewcommand\algorithmicensure{\textbf{Output:}}

\begin{algorithm}[t]
\small
\caption{KVCache-centric Scheduling Algorithm}\label{alg:kvcache_centric_scheduling}
\begin{algorithmic}[1]
\Require prefill instance pool $P$, decoding instance pool $D$, request $R$, cache block size $B$.
\Ensure the prefill and decoding instances $(p, d)$ to process $R$.
\State $\mathit{block\_keys} \leftarrow \text{PrefixHash}(R.\mathit{prompt\_tokens}, B)$
\State $\mathit{TTFT} \leftarrow \inf$
\State $p \leftarrow \emptyset$
\State $\mathit{best\_prefix\_len}, \mathit{best\_matched\_instance} \leftarrow \text{FindBestPrefixMatch}(P, \mathit{block\_keys})$
\For{$\mathit{instance} \in P$}
    \State $\mathit{prefix\_len} \leftarrow \mathit{instance.prefix\_len}$
    \State $\mathit{T_{queue}} \leftarrow \text{EstimatePrefillQueueTime}(\mathit{instance})$
    \If{$\frac{\mathit{best\_prefix\_len}}{\mathit{prefix\_len}} < {\bf kvcache\_balancing\_threshold}$}
        \Comment {Cache-aware prefill scheduling}
        \State $\mathit{T_{prefill}} \leftarrow \text{EstimatePrefillExecutionTime}(\text{len}(R.\mathit{prompt\_tokens}), \mathit{prefix\_len})$
        \If{$\mathit{TTFT} > \mathit{T_{queue}} + \mathit{T_{prefill}}$}
            \State $\mathit{TTFT} \leftarrow \mathit{T_{queue}} + \mathit{T_{prefill}}$
            \State $p \leftarrow \mathit{instance}$
        \EndIf
    \Else
        \Comment {Cache-aware and -balancing prefill scheduling}
        \State $\mathit{transfer\_len} \leftarrow \mathit{best\_prefix\_len} - \mathit{prefix\_len}$
        \State $\mathit{T_{transfer}} \leftarrow \text{EstimateKVCacheTransferTime}(\mathit{instance}, \mathit{best\_matched\_instance}, \mathit{transfer\_len})$
        \State $\mathit{T_{prefill}} \leftarrow \text{EstimatePrefillExecutionTime}(\text{len}(R.\mathit{prompt\_tokens}), \mathit{best\_prefix\_len})$
        \If{$\mathit{TTFT} > \mathit{T_{transfer}} + \mathit{T_{queue}} + \mathit{T_{prefill}}$}
            \State $\mathit{TTFT} \leftarrow \mathit{T_{transfer}} + \mathit{T_{queue}} + \mathit{T_{prefill}}$
            \State $p \leftarrow \mathit{instance}$
        \EndIf
    \EndIf
\EndFor
\State $d, \mathit{TBT} \leftarrow \text{SelectDecodingInstance}(D)$
\Comment {Load-balancing decoding scheduling}
\If{$\mathit{TTFT} > \mathit{TTFT\_SLO}$ \textbf{or} $\mathit{TBT} > \mathit{TBT\_SLO}$}
    \State \textbf{reject} $R$; \textbf{return}
\EndIf
\If{$\frac{\mathit{best\_prefix\_len}}{p.\mathit{prefix\_len}} > {\bf kvcache\_balancing\_threshold}$}
    \State $\text{TransferKVCache}(\mathit{best\_matched\_instance}, p)$
    \Comment {KVCache hot-spot migration}
\EndIf
\State \textbf{return} $(p, d)$
\end{algorithmic}
\end{algorithm}

\subsection{Prefill Global Scheduling}
\label{sec:prefill_global_scheduling}

Previous research on LLM serving typically uses a load-balancing strategy that evaluates the load on each instance based on the number of assigned requests. In Mooncake, however, the selection of prefill instances considers additional factors—not just load but also the prefix cache hit length and the distribution of reusable KVCache blocks. While there is a preference to route requests to prefill instances with longer prefix cache lengths to reduce computation costs, it may be beneficial to schedule them to other nodes to ensure overall system balance and meet TTFT SLOs. To address these complexities, we propose a cache-aware global scheduling algorithm that accounts for both the prefill time due to the prefix cache and the queuing time associated with the load on the instance.

Algorithm~\ref{alg:kvcache_centric_scheduling} details the mechanism for our cache-aware prefill scheduling. For every new request, its input tokens are divided into several blocks, and a hash key is computed for each block. This involves generating a hash key of tokens in a block concatenated with the hash key of the previous block (if available). The request's block keys are then compared one by one against each prefill instance's cache keys to identify the prefix match length ($prefix\_len$). Similar reuse logic is already implemented in vLLM, but the open-source version of vLLM only supports local KVCache caching.

With this matching information, Conductor estimates the corresponding execution time based on the request length and $prefix\_len$ (which varies by instance). It then adds the estimated waiting time for that request to get the TTFT on that instance. Finally, Conductor assigns the request to the instance with the shortest TTFT and updates the cache and queue times for that instance accordingly. If the SLO is not achievable, Conductor directly returns the HTTP 429 Too Many Requests response status code to the upper layers.

The backbone of this scheduling framework is straightforward, but complexities are hidden in the engineering implementation of various components. For example, to predict the computation time of the prefill stage for a request, we employ a predictive model derived from offline test data. This model estimates the prefill duration based on the request's length and prefix cache hit length. Thanks to the regular computation pattern of Transformers, the error bound of this prediction is small as long as enough offline data is available. The queuing time for a request is calculated by aggregating the prefill times of all queued requests. In practical implementations, TTFTs are computed in parallel, rendering the processing time negligible compared to the inference time.

More difficulty lies in predicting the transfer time because it is determined not only by the size of the transferred data but also by the current network status, especially whether the sending node is under congestion. This also necessitates the replication of hot KVCache blocks, which will be discussed in the next section.

\subsection{Cache Load Balancing}

In our \sys cluster, each prefill machine manages its own set of local prefix caches. The usage frequency of these caches varies significantly. For example, system prompts are accessed by almost every request, whereas caches storing content from a local long document may be used by only one user. As discussed in \S\ref{sec:prefill_global_scheduling}, Conductor's role is crucial in achieving an optimal balance between cache matching and instance load. Thus, from the perspective of the distributed cache system, load balancing also plays an important role. Specifically, it involves strategizing on how to back up caches to ensure that global prefill scheduling can achieve both high cache hits and low load.

A straw-man solution to this KVCache scheduling problem could be collecting the global usages of each block, using a prediction model to forecast their future usages, and making scheduling decisions accordingly. However, unlike the estimation of prefill time, workloads are highly dynamic and change significantly over time. Especially for a MaaS provider experiencing rapid growth in its user base, it is impossible to accurately predict future usage. Thus, we propose a heuristic-based automated hot-spot migration scheme to enhance cache load balancing.

\begin{wrapfigure}{r}{0.48\textwidth}
    \centering
    \includegraphics[width=0.47\textwidth]{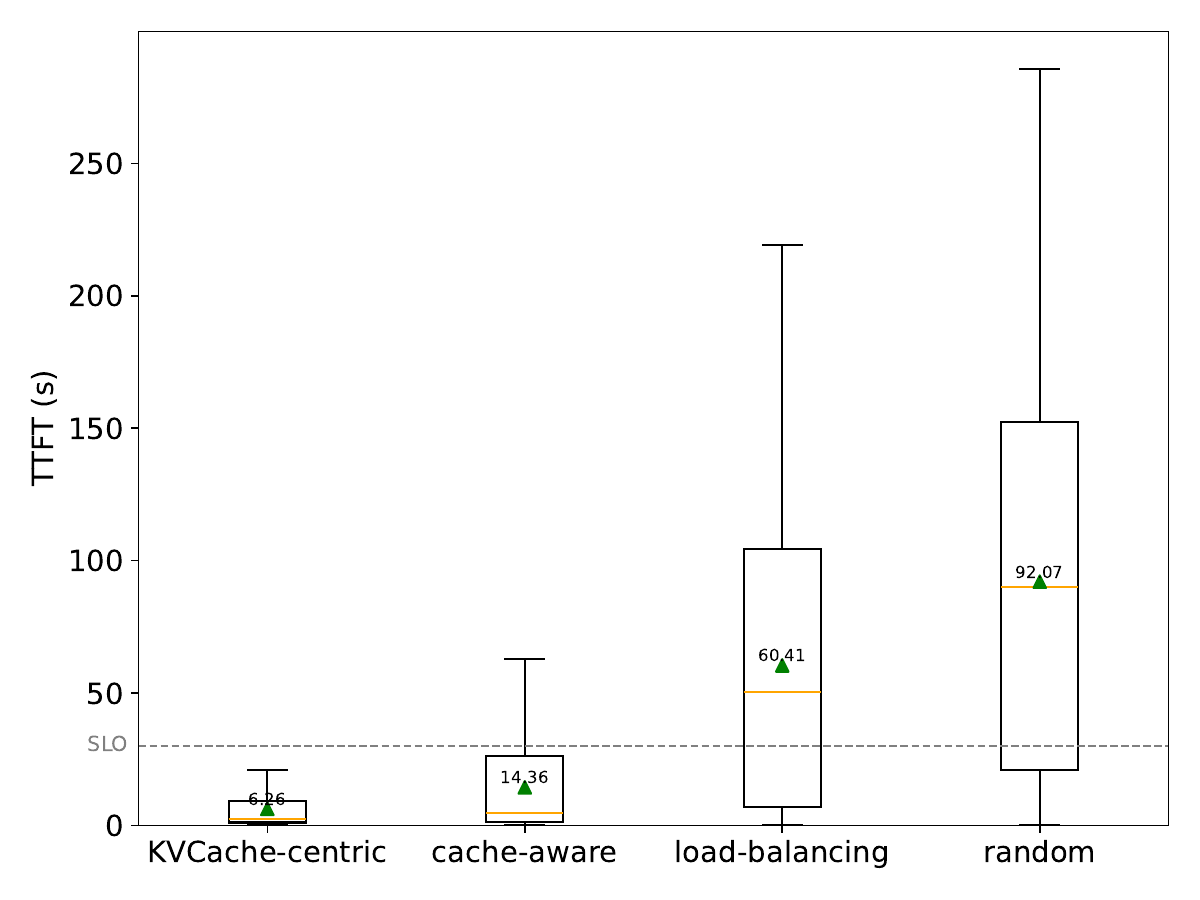}
    \caption{The prefill scheduling experiment in the Mooncake cluster.}
    \label{fig:prefill_global_scheduling}
    \end{wrapfigure}

As previously noted, requests may not always be directed to the prefill instance with the longest prefix cache length due to high instance load. In such cases, the conductor forwards the cache's location and the request to an alternative instance if the estimated additional prefill time is shorter than the transfer time. This instance proactively retrieves the KVCache from the holder and stores it locally. More importantly, we prefer to compute the input tokens if the best remote prefix match length is no larger than the current local reusable prefix multiplied by a threshold\footnote{This threshold is currently adjusted manually, but can be adaptively adjusted by an algorithm in the future.} Both strategies not only reduce the prefill time for requests but also facilitate the automatic replication of hot-spot caches, allowing for their broader distribution across multiple machines.

To validate the effectiveness of our strategy, we conducted a scheduling experiment that compares random scheduling and load-balancing scheduling with our strategy. We further compare the cache-aware scheduling described in \S\ref{sec:prefill_global_scheduling} and the KVCache-centric scheduling described in this section that considers cache load balancing. In random scheduling, a prefill instance is selected arbitrarily for each request. In load-balancing scheduling, the instance with the lightest load is chosen. To evaluate, we built a Mooncake cluster consisting of 8 prefill instances and 8 decoding instances, using idle machines overnight, and replayed 23,000 real-world requests for the experiment. We assessed the performance of each scheduling algorithm using the average TTFT and the TTFT SLO attainment rate. The experimental results, depicted in Figure~\ref{fig:prefill_global_scheduling}, demonstrate that both the cache-aware strategy and the cache load balancing strategy significantly reduce the TTFT of requests. Our KVCache-centric scheduling algorithm outperforms both random and load-balancing scheduling across both metrics. More experiment results can be found in \S\ref{sec:eval}.
\section{Overload-oriented Scheduling}\label{sec:oos}

Most existing work on LLM serving assumes that all requests will be processed, optimizing the throughput or the TTFT and TBT of requests accordingly. However, in real scenarios, processing every incoming request is neither economical nor realistic. For commercial inference services facing rapidly increasing volumes of user requests, the growth rate of the cluster's inference resources is far slower than the increase in incoming requests. As a result, overload is a common issue in current LLM serving, especially during peak times.

To balance costs and user experience, the system should process as many requests as possible until the system load reaches a predefined threshold. After this point, remaining requests will be either directly rejected or deferred for later retry. Mooncake, implemented as a disaggregated inference system, allows for more flexible scheduling strategies but also confronts unique scheduling challenges not present in non-disaggregated systems and not mentioned in previous works\cite{patel2023splitwise,zhong2024distserve,hu2024inference}.

In this section, we describe an early rejection policy designed specifically for a disaggregated architecture and address the load fluctuation caused by this approach. We then explore how predicting the generation length is necessary to mitigate these problems.

\subsection{Scheduling in Overload Scenarios}

In scenarios where system overload occurs, scheduling involves determining whether to accept or reject incoming requests based on the system load. A critical aspect of this process is defining what constitutes the ``system load'', as this definition influences the threshold at which requests are rejected. In conventional coupled systems, the prediction of TTFT and TBT can be complicated by interference between the prefill and decoding stages. Therefore, the load is often measured simply by the ratio of the number of requests being processed to the system's maximum capacity.

In contrast, Mooncake, with its disaggregated architecture, processes the prefill and decoding stages independently. Thus we use SLO satisfaction as a direct load measurement. Specifically, we define \( l_{ttft} \) and \( l_{tbt} \) as the TTFT and TBT SLO constraints for requests, respectively. The load for prefill and decoding instances is then determined by comparing the predicted maximum TTFT and TBT on an instance against \( l_{ttft} \) and \( l_{tbt} \). With these two criteria, Mooncake's scheduling requires two key decisions: first, whether to accept the prefill stage based on the prefill instance's load, and second, whether to proceed with the decoding stage depending on the decoding instance's load.

\subsection{Early Rejection}
\label{sec:early_rejection}

In practice, the individual load on prefill or decoding instances does not accurately reflect the actual number of requests processed by the system. This discrepancy arises due to a time lag between scheduling prefill and decoding instances for a single request. If a request is rejected by the decoding instance due to high load after the prefill stage has been completed, the computational resources expended during the prefill stage are wasted. Consequently, the actual number of successfully processed requests during prefill is less than that indicated by the load metric.

To address this issue, it is natural to advance the load assessment of the decoding instance to precede the beginning of the prefill stage. We refer to this strategy as {\bf Early Rejection}. Upon the arrival of a request, Conductor evaluates whether to accept the request based on the greater load between the prefill and decoding pools. Early Rejection significantly reduces ineffective computations from rejected requests and enhances load balancing.

\begin{figure*}[th]
\begin{center}
\includegraphics[width=\textwidth]{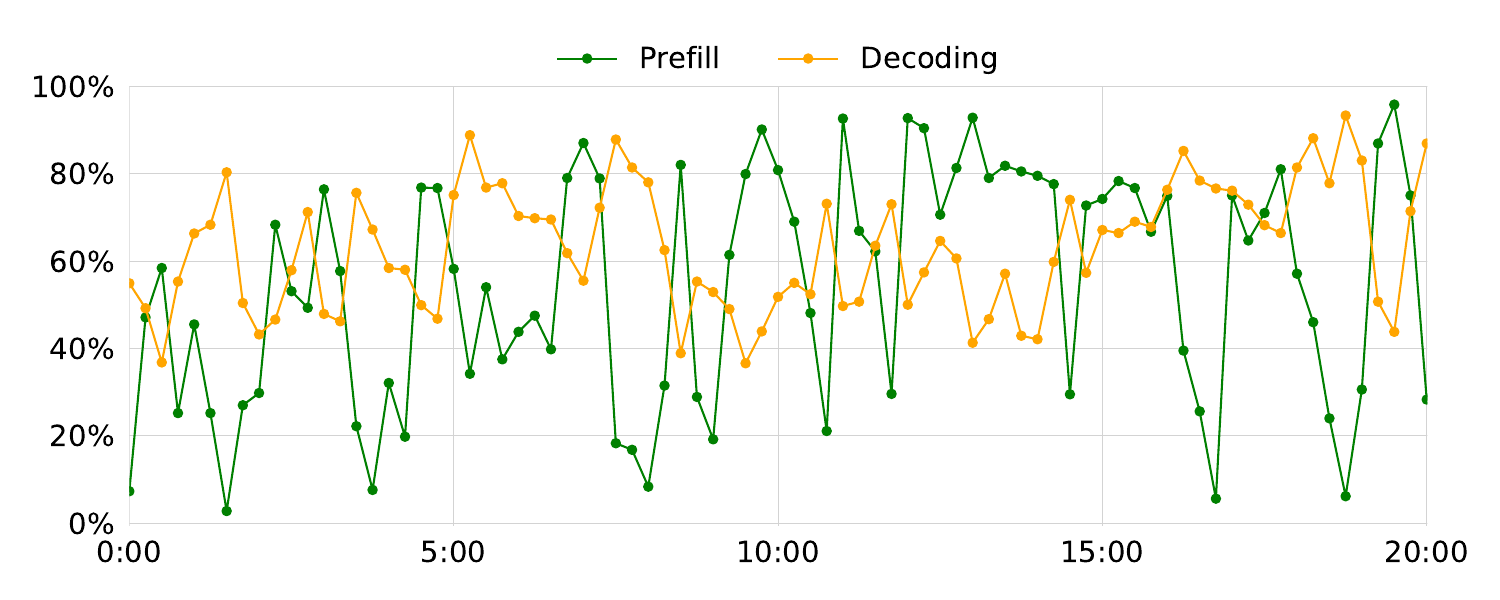}
\caption{The load of prefill and decoding instances over 20 minutes, before using the prediction-based early rejection.}
\label{fig:load_example}
\end{center}
\end{figure*}

\subsection{Load Fluctuation Caused by Early Rejection}

However, Early Rejection introduces new challenges. Figure~\ref{fig:load_example} shows the observed real-world instance load over a 20-minute period in a cluster of 20 machines after using the Early Rejection strategy. It highlights significant anti-phase fluctuations between prefill and decoding machines. This phenomenon becomes more pronounced in clusters with fewer prefill machines and in scenarios where the prefill stage takes longer.

\begin{figure*}[th]
\centering
\subcaptionbox{Early Rejection.\label{fig:early_rejection}}{%
    \includegraphics[width=0.85\textwidth]{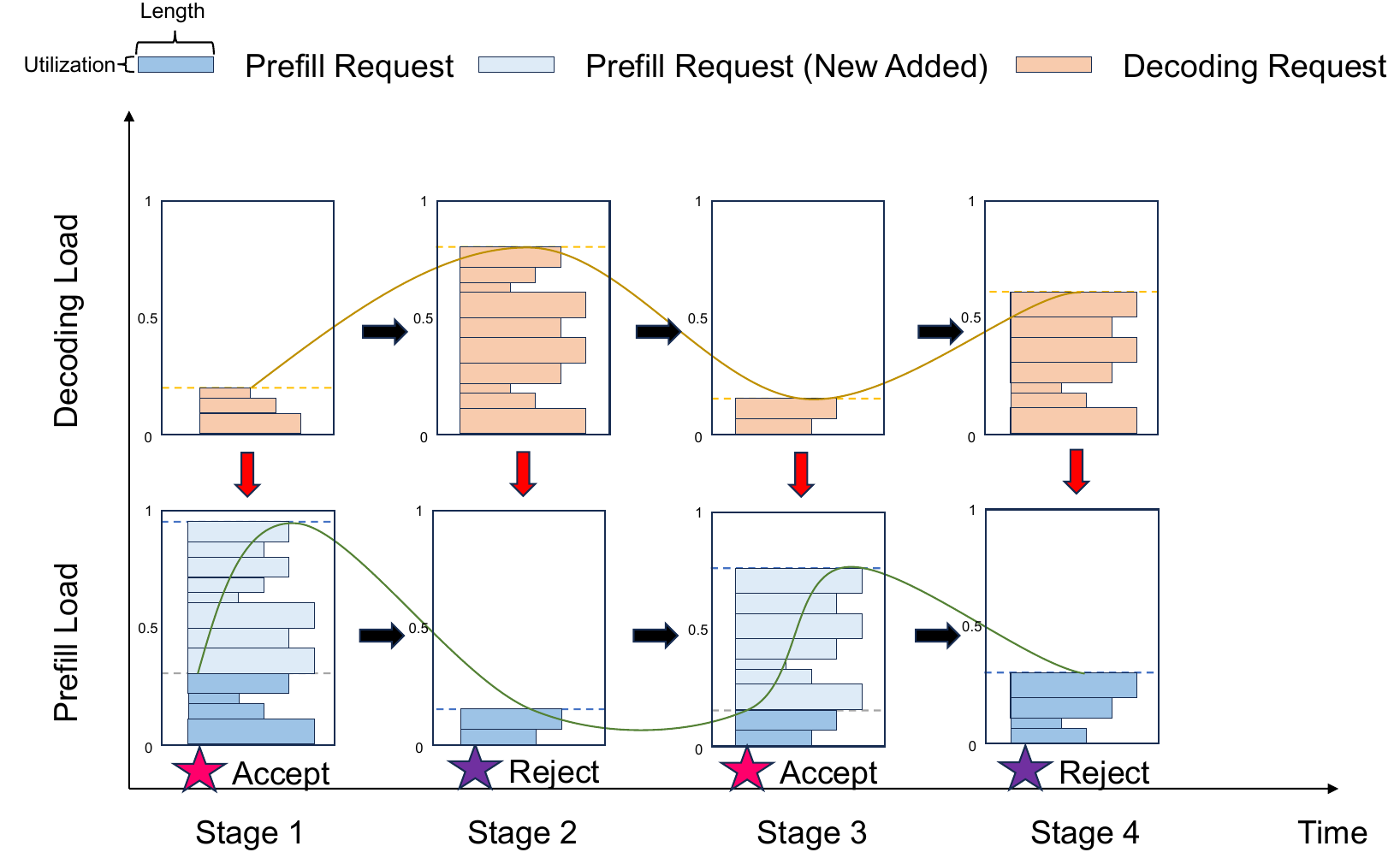}
}
\hspace*{0.04in}
\subcaptionbox{Early Rejection Based on Prediction.\label{fig:early_rejection_based_on_prediction}}{%
    \includegraphics[width=0.85\textwidth]{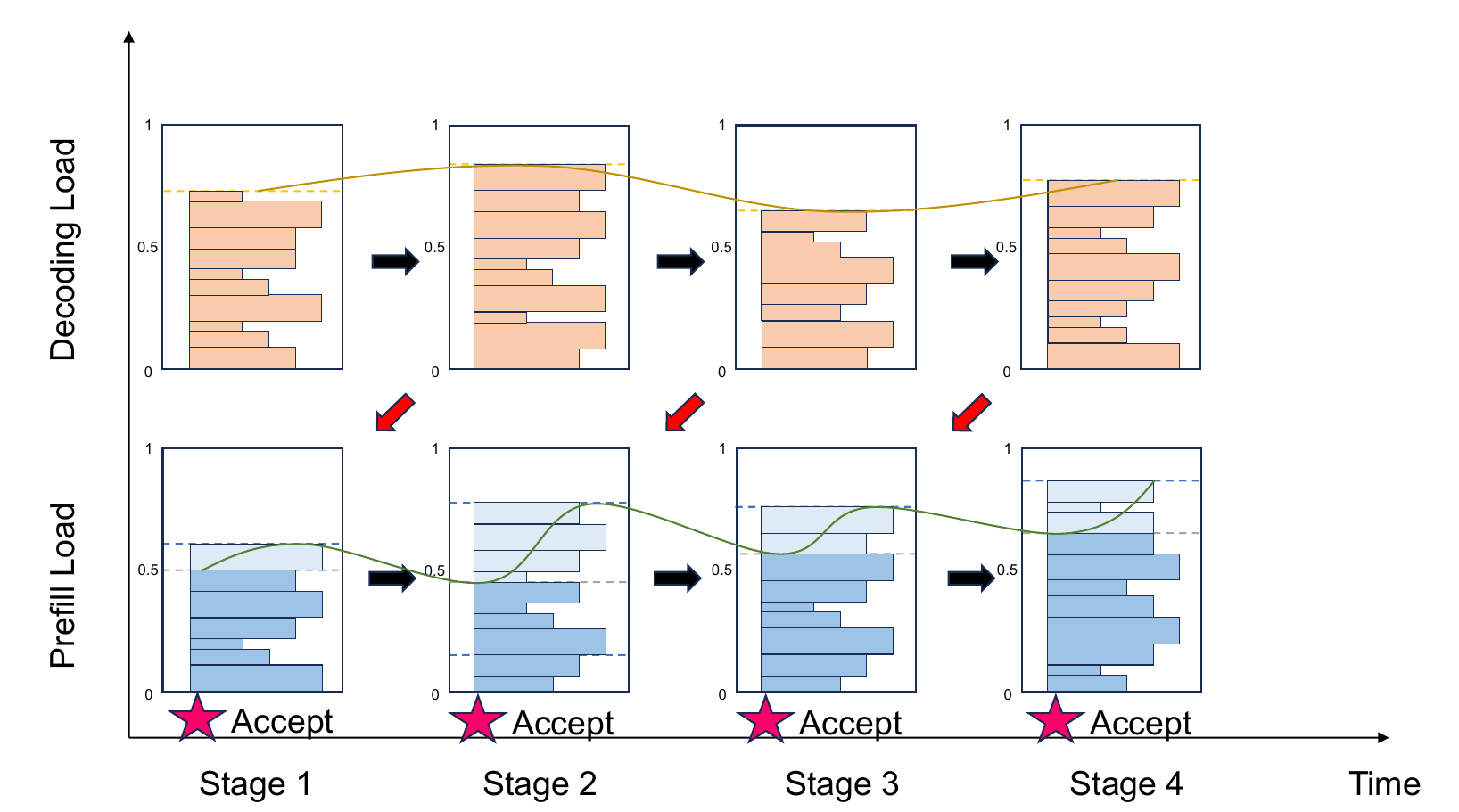}
}
\caption{Instance load when applying Early Rejection and Early Rejection Based on Prediction.}
\end{figure*}

Upon further exploration, we found that this load fluctuation problem is rooted in the time lag between predicting the decoding load and its actual execution. Scheduling based on the current decoding load is inherently delayed. This delay causes fluctuations and phase staggering between the loads on prefill and decoding instances, as illustrated in the theoretical example described in Figure~\ref{fig:early_rejection}. The green curve represents the load of prefill instances (scaled from 0 to 1), and the yellow curve represents the load of decoding instances.

In Stage 1, the load on both prefill and decoding instances is low, so Conductor accepts a large number of requests until the load on prefill instances reaches its limit. In Stage 2, requests processed by prefill instances are scheduled to decoding instances, causing the load on decoding instances to be high. Consequently, Conductor rejects incoming requests, leading to a lower load on prefill instances. In Stage 3, no new requests enter the decoding stage, resulting in a decreased load. At this point, Conductor again accepts a large number of requests until the prefill instances are fully loaded. In Stage 4, as the load on decoding instances increases, Conductor rejects requests, causing a low load on prefill instances. This severe fluctuation in load between prefill and decoding instances results in poor resource utilization of the inference cluster.

\subsection{Early Rejection Based on Prediction}
\label{sec:early_rejection_based_on_prediction}

To solve the load fluctuation problem, we propose a framework of Early Rejection Based on Prediction to address scheduling challenges in overload scenarios for disaggregated LLM serving systems like Mooncake. As illustrated in Figure~\ref{fig:early_rejection_based_on_prediction}, this framework predicts the decoding load after the prefill stage of incoming requests and uses this prediction to decide whether to accept the requests, which helps mitigate the fluctuation problem. The core component of this strategy is the accurate prediction of the decoding load for the subsequent period. We introduce two approaches for this:

\noindent\textbf{Request level:} Previous work highlights a significant challenge in predicting loads for LLM serving: the unknown output length of each request. If we could determine the output length in advance, it would be possible to estimate the TTFT and TBT much more accurately. This, in turn, would help predict the number of requests a decoding instance can complete and the number of new requests that will be added after a specified time, thereby obtaining the load at that time. However, predicting each request's output length is challenging due to high costs~\cite{hu2024inference} or low accuracy, especially under overload conditions where resources are scarce and accurate predictions are necessary, making request-level predictions particularly difficult.

\noindent\textbf{System level:} In contrast to request-level predictions, system-level predictions do not attempt to predict the completion time for individual requests. Instead, they estimate the overall batch count or the TBT status for instances after a specified time. This type of prediction is ongoing and requires less precision, making it more appropriate for overload scenarios.

In Mooncake, we currently utilize a system-level prediction strategy: we assume that each request's decoding stage takes a uniform time \( t_d \). First, for a given moment \( t \), requests that can be completed by the prefill instances at \( t \) are added to the uniform decoding instances. Next, requests that will be completed (i.e., their execution time exceeds \( t_d \)) before \( t \) are removed from the decoding instances. Finally, the average TBT ratio of all decoding instances to \( l_{tbt} \) is calculated to predict the load. The exploration of request-level prediction is left for future work.
\section{Evaluation}\label{sec:eval}

\subsection{End-to-end Performance}
\label{sec:end_to_end_performance}

\begin{table*}[th]
\begin{small}
\begin{center}
\caption{Datasets used in the end-to-end experiment.}
\label{tab:datasets}
\begin{tabular}{ccccc}
\toprule
Dataset & Avg Input Length & Avg Output Length & Cache Ratio & Arrival Pattern \\
\midrule
ArXiv Summarization~\cite{cohan2018discourse} & 8088 & 229 & \textasciitilde 0\% & Poisson Process \\
L-Eval~\cite{an2023leval} & 19019 & 72 & >80\% & Poisson Process \\
Simulated Data & 16k, 32k, 64k, 128k & 512 & 50\% & Poisson Process \\
Real Data & 7955 & 194 & \textasciitilde 50\% & Timestamp-based \\
\bottomrule
\end{tabular}
\end{center}
\end{small}
\end{table*}

This section evaluates the end-to-end performance of Mooncake under different datasets and various workloads. 
As stated before, to protect proprietary information and facilitate reproducibility, all the experimental results reported in this paper are based on a dummy model that follows the same architecture as LLaMA2-70B.

\textbf{Testbed}\quad During the experiments, the system was deployed on a high-performance computing node cluster to test performance. Each node in the cluster is configured as follows: 8 NVIDIA-A800-SXM4-80GB GPUs, each with 80GB HBM, connected by NVLINK; equipped with RDMA network cards that support up to 800 Gbps of interconnect bandwidth between nodes. Each node deploys either a prefill instance or a decoding instance according to the startup parameter.

\textbf{Dataset and Workload}\quad Building upon previous research~\cite{agrawal2024taming, zhong2024distserve, wu2024loongserve}, we selected or designed the datasets as outlined in Table~\ref{tab:datasets}. In addition to utilizing public datasets, we generated a batch of simulated data featuring predefined lengths and prefix cache ratios for our experiments. To examine performance in real-world scenarios, we constructed a dataset consisting of 23,000 real request traces, each annotated with an arrival timestamp. Experiments involving real request traces were conducted by replaying these requests according to their actual arrival times. For other scenarios, we simulated requests using a Poisson arrival process and controlled the request rate through RPS (Requests per Second).

\textbf{Metric}\quad  In the experiments, we focus on the throughput performance of various systems under defined SLOs. We measure the TTFT and TBT across different RPS rates, where a higher RPS signifies improved throughput. To assess whether the majority of requests satisfy the SLOs, we use the 90th percentile (P90) values of TTFT and TBT as the ultimate metrics. As mentioned in \S\ref{sec:preliminary}, the thresholds for TTFT and TBT are set by multiplying the lowest observed RPS values by factors of 10 and 5, respectively. Exceeding these thresholds indicates a failure to meet the SLOs, and the corresponding consumed resources are considered as wasted. For ease of comparison, we normalize all TTFT and TBT values against these upper limits, establishing a baseline of 1.0.

\textbf{Baseline}\quad We employ vLLM, one of the state-of-the-art open-source LLM serving systems, as our experimental baseline. vLLM incorporates continuous batching and PagedAttention technologies, significantly boosting inference throughput. Despite its strengths, vLLM's design, which couples the prefill and decoding stages of inference requests, can cause disruptions during decoding in scenarios involving long contexts.

\begin{figure*}[th]
\includegraphics[width=\linewidth]{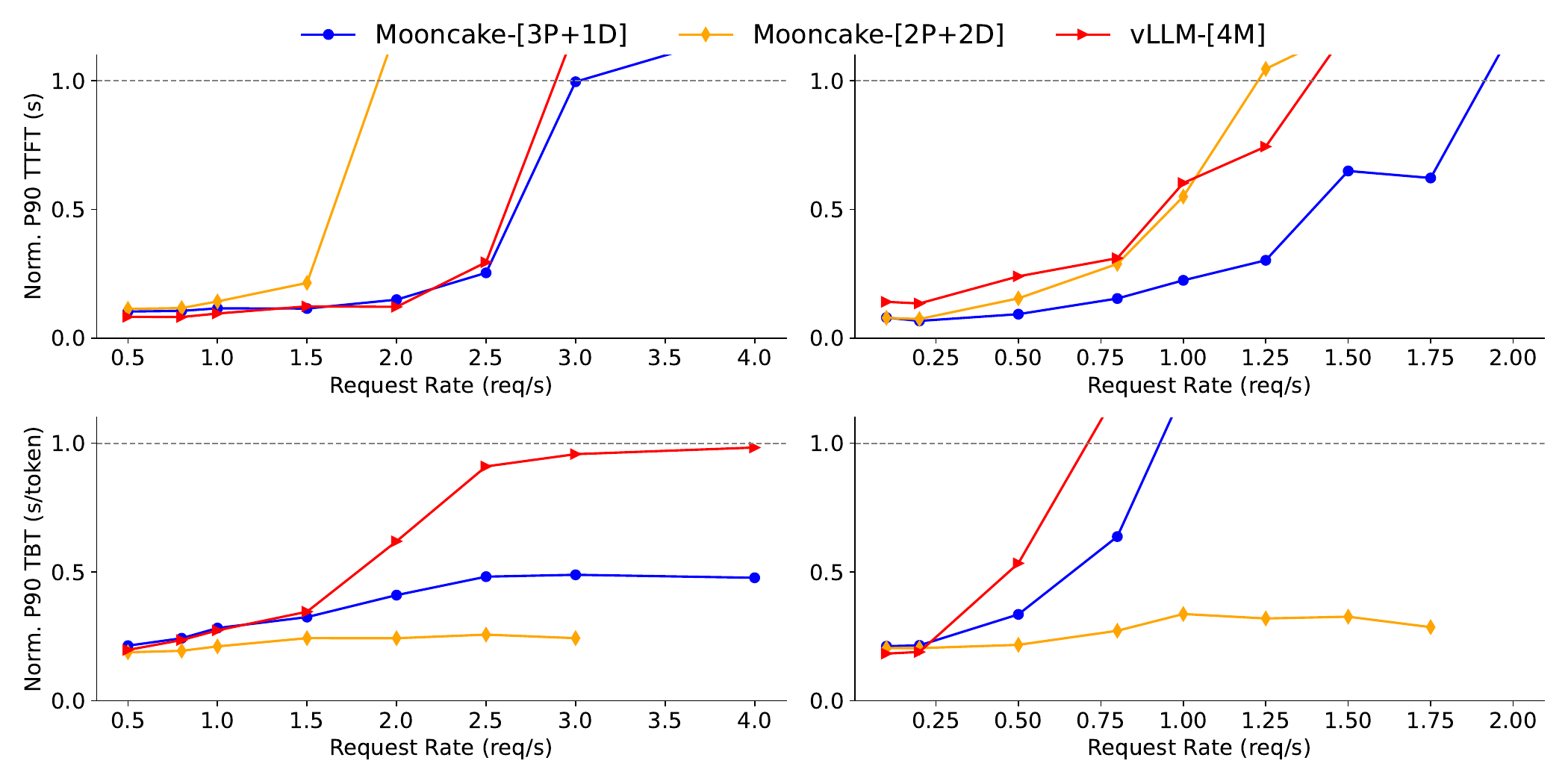}
\vspace{-0.05in}
\hspace*{1.0in}{ArXiv Summarization}\hspace*{1.9in}{L-Eval}
\vspace{0.1in}
\caption{End-to-end experiments of \sys and vLLM on the ArXiv Summarization and L-Eval datasets}
\label{fig:public_dataset_e2e}
\end{figure*}

\subsubsection{Public Datasets}
\label{sec:public_dataset_e2e}

This section evaluates the performance of Mooncake and vLLM in end-to-end tests on public datasets using ArXiv Summarization and L-Eval. We establish a baseline using a cluster of four vLLM instances, denoted as vLLM-[4M]. In contrast, Mooncake is configured in two distinct setups: one cluster consists of three prefill instances and one decoding instance, labeled Mooncake-[3P+1D], and the other has two prefill and two decoding instances, labeled Mooncake-[2P+2D]. The results, depicted in Figure~\ref{fig:public_dataset_e2e}, demonstrate that on the ArXiv Summarization and L-Eval datasets, Mooncake-[3P+1D] achieves throughput improvements of 20\% and 40\%, respectively, over vLLM-[4M] while satisfying SLOs. Moreover, Mooncake's throughput on the L-Eval dataset is further enhanced by prefix caching, which significantly reduces prefill time. However, despite having lower TBT latency, Mooncake-[2P+2D] does not perform as well on the TTFT metric compared to Mooncake-[3P+1D] and vLLM-[4M]. This discrepancy arises from an imbalance in the load between prefill and decoding instances. In real-world clusters, the demand for prefill and decoding instances generally remains stable over certain periods, with only minor temporary imbalances. Thus, the proportion of prefill and decoding instances can be preset. Future research will explore more flexible deployment and conversion methods.

\begin{figure*}[th]
\includegraphics[width=\linewidth]{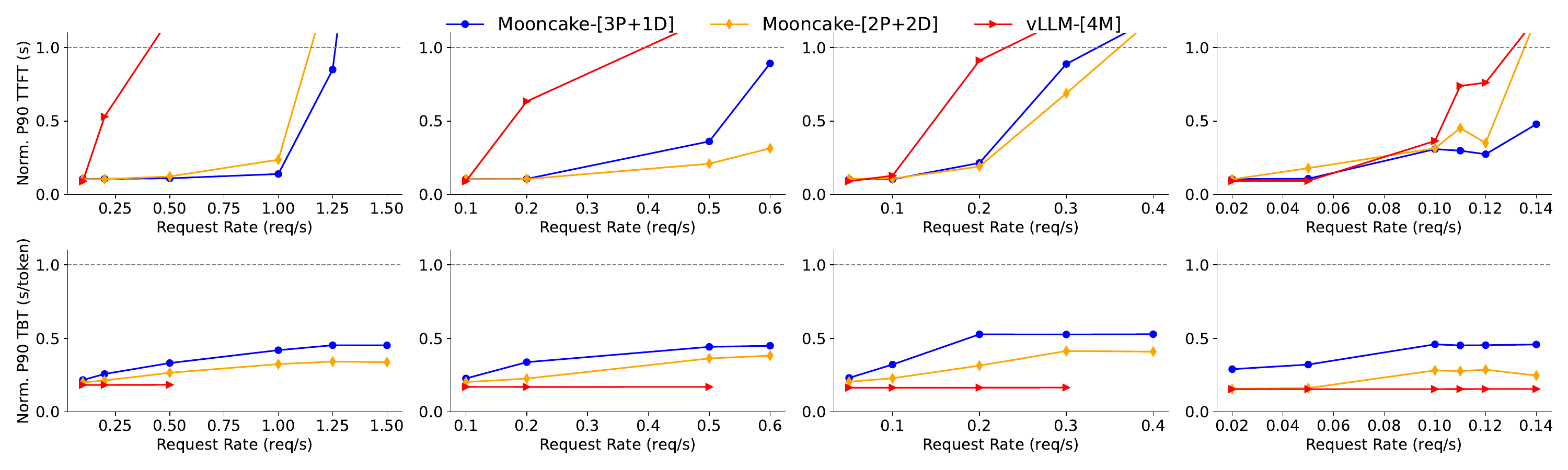}
\vspace{-0.05in}
\hspace*{0.5in}{16k prompt}\hspace*{0.7in}{32k prompt}\hspace*{0.7in}{64k prompt}\hspace*{0.7in}{128k prompt}
\vspace{0.1in}
\caption{End-to-end experiments of \sys and vLLM on simulated data.}
\label{fig:mock_dataset_e2e}
\end{figure*}

\subsubsection{Simulated Data}

In this section, we employ simulated data for an end-to-end experiment. The cluster configuration is the same as in \S\ref{sec:public_dataset_e2e}, utilizing Mooncake configurations of [3P+1D], [2P+2D], and vLLM-[4M]. Notably, the long-context requests in simulated data significantly disrupt the decoding stage of vLLM. To counteract this, vLLM processes requests individually, rather than in batches. The results of the experiment are presented in Figure~\ref{fig:mock_dataset_e2e}. Although Mooncake employs batch processing, its two-stage disaggregation design effectively minimizes the impact of the prefill stage on the decoding stage, ensuring it never breaks the TBT SLO. Mooncake demonstrates significantly higher throughput, with enhancements ranging from 50\% to 525\%, while adhering to the same TTFT and TBT SLO constraints compared to vLLM.

\begin{figure*}[th]
\includegraphics[width=\linewidth]{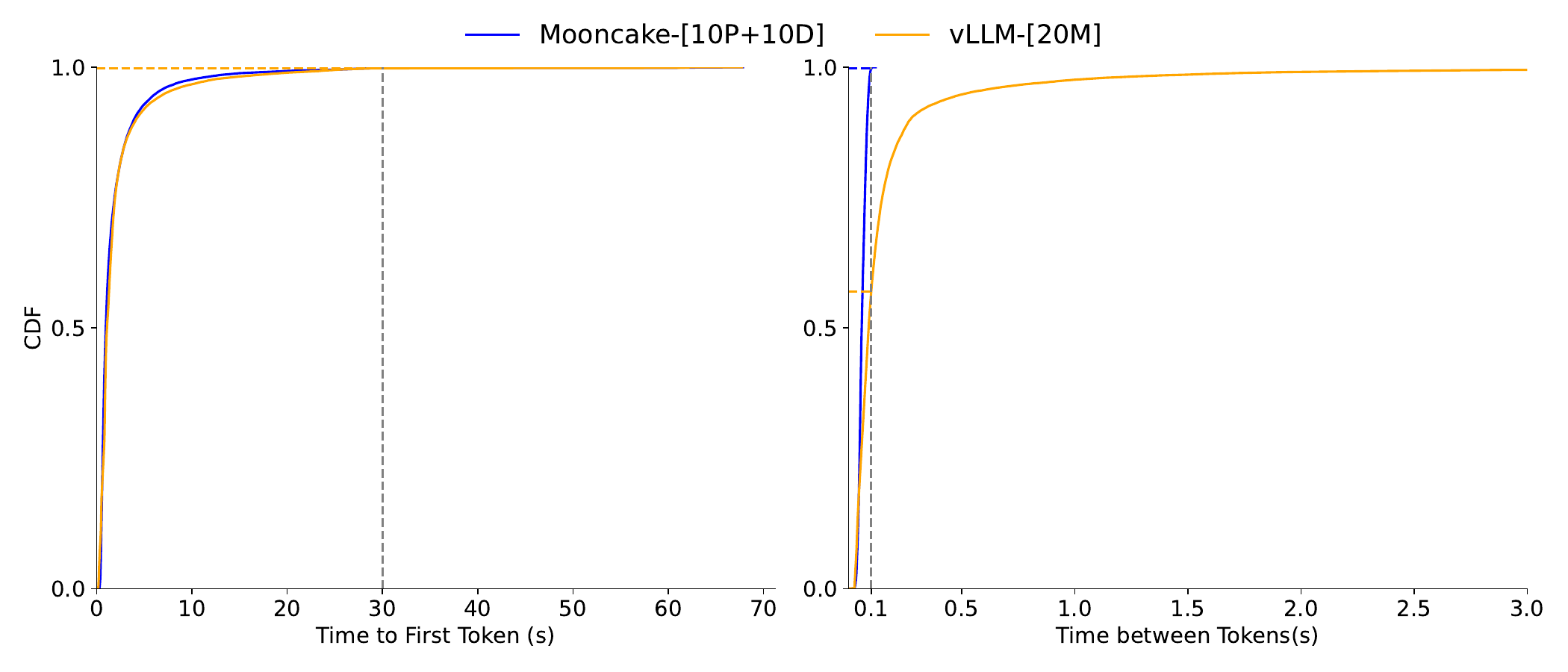}
\caption{Request TTFT and TBT distributions of \sys and vLLM under real workloads}
\label{fig:real_data_e2e}
\end{figure*}

\subsubsection{Real Workload}
\label{sec:real_data_e2e}

We further utilize 10 prefill instances and 10 decoding instances, labeled Mooncake-[10P+10D], along with 20 instances of vLLM, referred to as vLLM-[20M], to replay {\bf real request traces} and conduct load tests on both Mooncake and vLLM. In this experimental setup, the upper limit for the TTFT is set at 30 seconds, while the TBT threshold is capped at 0.1 seconds per token. Figure~\ref{fig:real_data_e2e} presents the CDF (Cumulative Distribution Function) plots for the TTFT and TBT for the two systems. The TTFT distributions for both Mooncake-[10P+10D] and vLLM-[20M] are nearly identical, with almost 100\% of requests meeting the TTFT SLO. However, while approximately 100\% of the requests for Mooncake-[10P+10D] satisfy the TBT SLO, only 57\% of the requests for vLLM-[20M] meet this criterion, with some requests exhibiting extremely high TBTs. In this experiment, Mooncake can process approximately 75\% more requests while adhering to the SLOs.

\subsection{Performance in Overload Scenarios}

In this section, we evaluate performance under overload scenarios, focusing on the maximum number of requests the system can handle, as discussed in \S\ref{sec:oos}. The baseline strategy, which rejects requests based on load before both stages start, leads to resource wastage by rejecting requests already processed in the prefill stage. In contrast, we propose the Early Rejection and Early Rejection based on Prediction strategies, detailed in \S\ref{sec:early_rejection} and \S\ref{sec:early_rejection_based_on_prediction}, respectively. These strategies take the system's load into comprehensive consideration, and hence reduce unnecessary request rejections.

Specifically, we built a Mooncake cluster with 8 prefill instances and 8 decoding instances and tested it using real traces from 23,000 requests. To simulate overload scenarios, we increased the replay speed to 2x.

\begin{table*}[th]
\begin{small}
\begin{center}
\caption{Number of requests rejected by the system under the overloaded-scenario experiment.}
\label{tab:ooc_result}
\begin{tabular}{cccc}
\toprule
 & Baseline & Early Rejection & Early Rejection based on Prediction \\
\midrule
Number of rejected requests & 4183 & 3771 & 3589 \\
\bottomrule
\end{tabular}
\end{center}
\end{small}
\end{table*}

Table~\ref{tab:ooc_result} shows Mooncake's performance under different strategies. With the baseline strategy, the system rejects 4,183 requests. In contrast, under the Early Rejection and Early Rejection based on Prediction strategies, Mooncake rejects 3,771 and 3,589 requests, respectively. This demonstrates that by rejecting requests early, Mooncake can avoid unnecessary prefill computations, thereby improving the effective utilization of system resources. Furthermore, by predicting the load of decoding instances, Mooncake can mitigate load fluctuations, increasing the request handling capacity.

\section{Related Work}

Significant efforts have been dedicated to enhancing the efficiency of LLM serving systems through scheduling, memory management, and resource optimization. 
Production-grade systems like FasterTransformer~\cite{fastertransfomer}, TensorRT-LLM~\cite{tensorrtllm}, and DeepSpeed Inference~\cite{aminabadi2022deepspeed} are designed to significantly boost throughput. 
Orca~\cite{yu2022orca} employs iteration-level scheduling to facilitate concurrent processing at various stages, while vLLM~\cite{kwon2023efficient} leverages dynamic KVCache management to optimize memory. 
FlexGen~\cite{sheng2023flexgen}, SARATHI~\cite{agrawal2024taming}, and FastServe~\cite{wu2023fast} incorporate innovative scheduling and swapping strategies to distribute workloads effectively across limited hardware, often complementing each other's optimizations.

Our design of Mooncake builds on these developments, particularly drawing from the open-source community of vLLM, for which we are deeply appreciative.

Moreover, recent research shares our insight into separating the prefill and decoding stages, leading to a disaggregated architecture that enhances system throughput. The arXiv publication of Splitwise~\cite{patel2023splitwise} is at the early stage of the development of Mooncake, which further motivated our progress. Many concurrent works corroborate our findings, including DistServe~\cite{zhong2024distserve}, which optimizes resource allocation and parallel strategies for each stage to maximize GPU goodput, and TetriInfer~\cite{hu2024inference}, which incorporates both chunked prefill and two-stage disaggregation along with a predictive two-stage scheduling algorithm to optimize resource utilization.

Prefix caching is also widely adopted to enable the reuse of KVCache across multiple requests, reducing computational overhead in LLM inference systems~\cite{tensorrtllm, kwon2023efficient}. Prompt Cache~\cite{gim2023prompt} precomputes and stores frequently used text KVCache on inference servers, facilitating their reuse and significantly reducing inference latency. SGLang~\cite{zheng2023efficiently} leverages RadixAttention, which uses a least recently used (LRU) cache within a radix tree structure to efficiently enable automatic sharing across various reuse patterns.

Among these approaches, AttentionStore~\cite{gao2024attentionstore}, a concurrent work with us, proposes a hierarchical KVCache system that utilizes cost-effective memory and storage media to accommodate KVCache for all requests. The architecture of Mooncake shares many design choices with AttentionStore. However, in long-context inference, the KVCache becomes extremely large, requiring high capacity and efficient data transfer along with KVCache-centric global scheduling. 
Additionally, Mooncake is not a standalone cache service; it incorporates both a memory-efficient cache storage mechanism and a cache-aware scheduling strategy, further improving prefix caching efficiency.

Furthermore, recent research~\cite{srivatsa2024preble} has started exploring the scheduling of prompts, which is essentially KVCache-centric scheduling. We corroborate many results in this area, although the real reusability in our online traces is much smaller than the results reproduced by open-source benchmarks. Theoretically, up to only 55\% of the KVCache can be reused in our current workloads, even if we assume both the capacity of storage and the TTFT SLO are infinite. However, this reusability highly depends on the application scenario and can be as large as 90\% for certain scenarios, such as our chat-to-paper service \url{https://papers.cool/}. We also emphasize the need for overload-oriented scheduling subject to SLOs, rather than merely throughput-oriented scheduling.

\section{Future Work}

Disaggregating different parts of LLM serving into dedicated resource pools is key to Mooncake's high resource utilization. 
In the future, we plan to explore more opportunities along this path, particularly the potential use of heterogeneous accelerators. 
Current flagship accelerators balance multiple metrics such as computational power, memory bandwidth, and capacity, making them versatile but not optimal in every single metric. 
For instance, considering only bandwidth per dollar or bandwidth per watt, current GDDR and even LPDDR solutions can be an order of magnitude better than flagship accelerators.
We are also particularly interested in new technologies that use process-in-memory~\cite{10.1145/3400302.3415640,9474146,10213232,10155455} or hybrid bonding~\cite{10.1145/3639038,ddrasm,10.1145/2541228.2541231,7174528,7551408} techniques to implement memory-oriented devices that could offer both high bandwidth and high capacity in the near future. 
These technologies would be ideal for reducing the cost of executing memory-bound operations in the decoding phase.

Furthermore, in a heterogeneous accelerator environment that includes both computation-oriented and bandwidth-oriented accelerators, we can explore more advanced disaggregation architectures. 
For example, unlike other linear transformation operators, the arithmetic intensity of the attention operator in the decoding phase is only proportional to the number of attention heads divided by the number of key/value heads. 
This intensity cannot be increased by increasing the batch size and is typically more memory-bound than other operators. 
Therefore, it is possible to separate the attention operator from other linear operators to improve resource utilization further. 
According to our preliminary simulated results~\cite{chen2024efficient}, this architecture has great potential to increase overall throughput. 
Additionally, the recently proposed MLA operator by DeepSeek-v2~\cite{deepseekai2024deepseekv2} directly increases arithmetic intensity, solving this problem from another angle and showing great promise.

As an orthogonal direction, many algorithms aim to reduce the size of KVCache, benefiting Mooncake in two important ways: 1) increasing the batch size for better utilization and 2) improving the KVCache cache hit ratio to reduce prefill costs. 
This is currently a very active area, including different methods for compressing KVCache~\cite{yu2024effectively,cai2024pyramidkv,2.2.28167.37282,he2024zipcache,liu2024intactkv,liu2024minicache}, selecting important tokens by various metrics~\cite{guo2024attention,devoto2024simple,yao2024sirllm,yang2024pyramidinfer,adnan2024keyformer,li2024snapkv,zhang2023h2o}, sharing KVCache across different layers~\cite{zuhri2024mlkv,wu2024layercondensed,ccai}, or using hybrid architectures with operators that do not use KVCache~\cite{sun2024cache,gu2024mamba,peng2023rwkv,dao2024transformers,lieber2024jamba,botev2024recurrentgemma}.

In terms of scheduling, we are developing an advanced policy that accounts for varying request priorities and scenarios with different TTFT/TBT SLOs. 
This policy is designed to enhance the responsiveness and efficiency of our system under diverse operational conditions. 
Effective management of KVCache, including replication, migration, and specialized eviction policies for partial hits and expiration scenarios, is also crucial for optimizing cache reuse.
Additionally, we plan to dynamically balance prefill and decoding instances and investigate strategies for utilizing idle resources through batch-oriented offloading tasks. 
This approach will allow us to maximize resource utilization during fluctuating workloads.
\section{Conclusion}

This paper presents Mooncake, a KVCache-centric disaggregated architecture designed for efficiently serving LLMs, particularly in handling long contexts and overloaded scenarios. We discuss the necessity, challenges, and design choices involved in balancing the goal of maximizing overall effective throughput while meeting latency-related SLO requirements.

\bibliographystyle{unsrt}
\bibliography{main.bib}

\end{document}